\RequirePackage{lineno}
\documentclass[superscriptaddress,aps,preprint,amsmath,amssymb,floatfix]{revtex4-1}

\usepackage{graphicx,morefloats}
\usepackage{subfigure} 
\usepackage{color,soul}
\usepackage{bm}
\usepackage{amsmath}
\usepackage{amssymb}
\usepackage{times}
\usepackage{hyperref}
\usepackage{dsfont}
\usepackage{bbm}

\usepackage{gensymb}

\newcommand{\be}[0]{\begin{equation}}
\newcommand{\ee}[0]{\end{equation}}
\newcommand{\ba}[0]{\begin{eqnarray}}
\newcommand{\ea}[0]{\end{eqnarray}}
\newcommand{\mx}[0]{\begin{pmatrix}}
\newcommand{\ex}[0]{\end{pmatrix}}

\begin{document}

\hyphenation{va-ni-sh-ing}

\begin{center}

\thispagestyle{empty}

{\large\bf Topological Semimetal Driven by Strong Correlations and Crystalline Symmetry}
\\
[0.3cm]

Lei\ Chen$^{1}$,
Chandan Setty$^{1}$,  Haoyu\ Hu$^1$, Maia\ G.\ Vergniory$^{2,3}$, Sarah E. Grefe$^4$,
Lukas\ Fischer$^5$, Xinlin\ Yan$^5$, Gaku Eguchi$^5$,
Andrey\  Prokofiev$^5$, Silke\  Paschen$^{5,1,\ast}$, Jennifer Cano$^{6,7,\ast}$, and Qimiao Si$^{1,\ast}$
\\[0.3cm]

$^1$Department of Physics and Astronomy, Rice Center for Quantum Materials, Rice University, Houston, Texas 77005, USA\\[-0.cm]

$^2$Max  Planck  Institute  for  Chemical  Physics  of  Solids,  01309  Dresden,  Germany\\[-0.cm]

$^3$Donostia International  Physics  Center,  P. Manuel  de Lardizabal 4,  20018 Donostia-San Sebastian,  Spain\\[-0.cm]

$^4$Theoretical Division, Los Alamos National Laboratory, Los Alamos, New Mexico 87545, USA\\[-0.0cm]

$^5$Institute of Solid State Physics, Vienna University of Technology, Wiedner Hauptstr. 8-10, 1040
Vienna, Austria\\[-0.0cm]

$^6$Department of Physics and Astronomy, Stony Brook University, Stony Brook, NY 11794, USA\\[-0.cm]

$^7$Center for Computational Quantum Physics, Flatiron Institute, New York, NY 10010, USA

\end{center}

\vspace{0.16cm}
{\bf 
Electron correlations amplify quantum fluctuations and, as such, they have been recognized 
as the origin of  a rich landscape of quantum phases.
Whether and how they lead to gapless topological states is an outstanding question, and 
 a framework that allows for determining novel phases and identifying new materials is in pressing need.
Here we advance a general approach, in which strong correlations \cite{Kei17.1,Pas21.1} cooperate
with crystalline symmetry \cite{Armitage2017,cano2021band} to drive gapless topological states. 
We test this materials
design principle by exploring Kondo lattice models and materials 
whose space group symmetries may promote different kinds of electronic degeneracies, 
with a particular focus on square-net systems. Weyl-Kondo nodal-line semimetals -- with nodes pinned to the Fermi energy -- are identified. We describe how this approach can be applied to
discover strongly correlated topological semimetals, identify
three heavy fermion compounds as new candidates, provide first direct experimental evidence for our prediction in Ce$_2$Au$_3$In$_5$, and discuss how our approach may lead to many more.
Our findings illustrate the potential of the proposed materials
design principle to guide  the search for new topological 
metals in a broad range of  strongly correlated systems.
}
\vspace{0.6cm}
\noindent E-mails: qmsi@rice.edu; jennifer.cano@stonybrook.edu; paschen@ifp.tuwien.ac.at

\newpage
Electron correlations and topology are well established as
engines for surprising and potentially functional properties.
Strong correlations promote quantum fluctuations, which engender abundant phases of matter 
and 
various quantum phase transitions \cite{Kei17.1,Pas21.1}.
Meanwhile, extensive 
 developments have taken place 
in noninteracting electron systems, especially those with sizable spin-orbit couplings (SOC).
In particular, the role of space group symmetry in determining and classifying
symmetry-protected topological phases has been
highlighted \cite{Bradlyn2017,Cano2018,Po2017,Watanabe2017,cano2021band}.
In the ``hydrogen atom" version of topological 
semimetals in two and three dimensions (2D and 3D), space-group
symmetry enforces Dirac nodes in the
honeycomb lattice -- as realized in graphene \cite{CastroNeto09} -- 
and diamond lattice \cite{Fu2007}, respectively. They set the stage for a systematic search of weakly correlated topological materials
in terms of the constraints of
space group symmetries
on noninteracting bandstructures for a variety of crystal structures,
which has led to a large topological materials database \cite{Vergniory2019,Zhang2019}.

We can expect that the intersection of these two
fields will be especially fertile in breeding novel quantum 
phases, but it remains a largely open terrain \cite{Maciejko15,Rachel2010,Dzero2010}
especially for the case with gapless bulk excitations \cite{Pas21.1,Sch16.2}.
The fractional quantum Hall effect amply demonstrates the capability of strong correlations to drive
gapped topological states. By contrast, correlated gapless electronic topology has been much less explored.
The usual approach starts from noninteracting symmetry-protected topological states. 
It considers the interactions as a perturbation 
or as producing a symmetry-breaking ordered state -- such as a magnetic order -- that
in turn modifies the weakly-correlated topology \cite{Armitage2017}.
In the opposite limit, where the electrons' correlations dominate over their kinetic energy,
there has been a considerable lack of
strongly correlated gapless topological materials.
It is pressing to realize such materials,
which are important in their own right and may also serve as anchoring points to explore the 
overall landscape of strongly correlated gapless topology.
Recently, a non-perturbative study 
has led to a Weyl-Kondo semimetal phase \cite{Lai2018,Grefe2020_2,Grefe2021}
in a toy model associated with the diamond lattice.
Concurrently, such a phase was advanced experimentally in a 
 cubic heavy fermion 
compound Ce$_3$Bi$_4$Pd$_3$  
based on measurements of the specific heat and a spontaneous Hall effect
\cite{Dzsaber2017,Dzsaber2021,Dzsaber2019.x}.
The interplay between significant electron correlations and lattice geometry is also being explored 
experimentally in other  materials such as the
 kagome metals \cite{Asaba2021,Kan20.2,Yao18.1x}.

\begin{figure}[t]
\includegraphics[width=0.9\columnwidth]{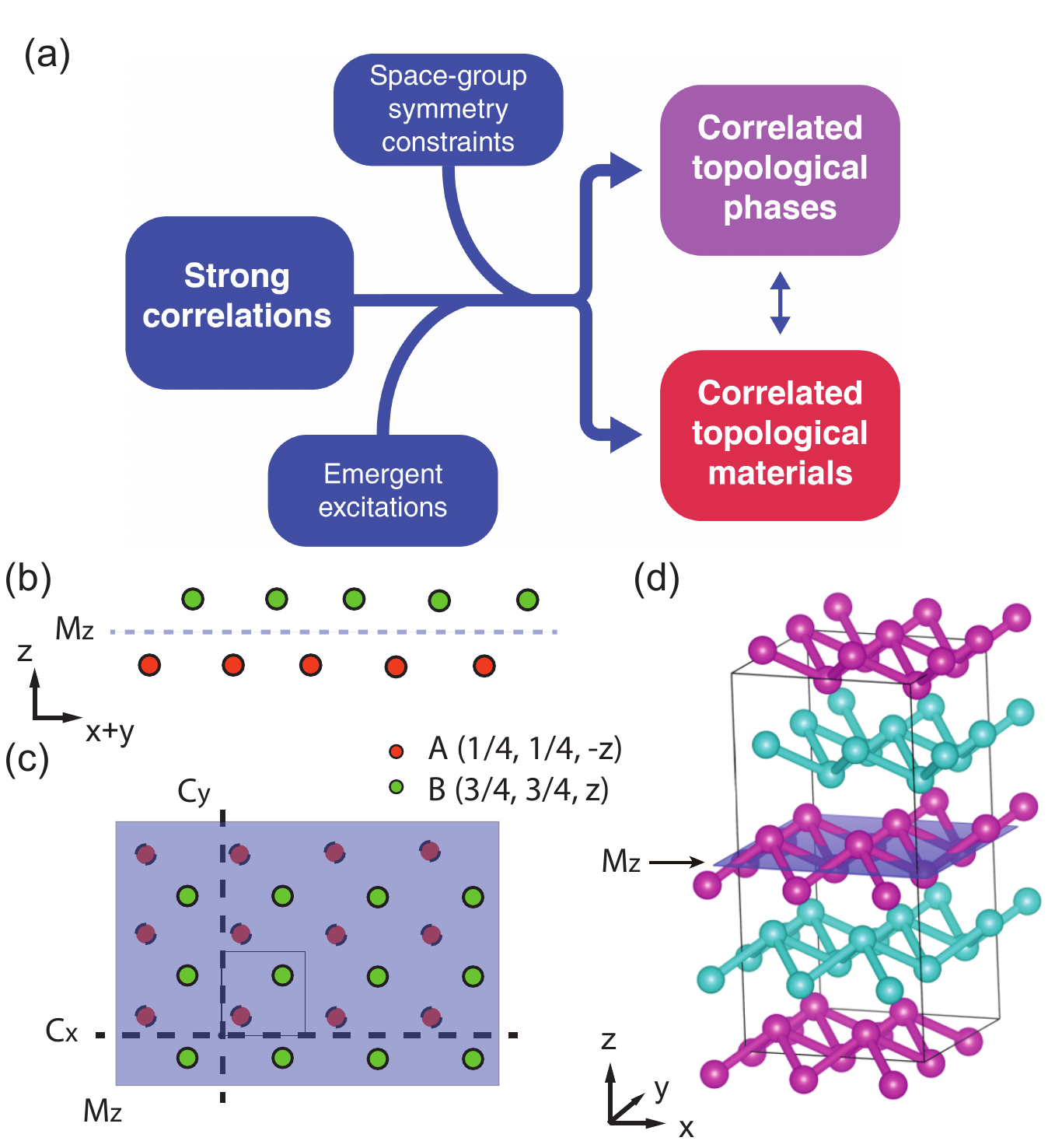}
\vskip 0.15 in
\caption{ {\bf Materials design principle and space-group symmetry.}
{\bf a,} The proposed cooperation between strong correlations and space-group symmetry in realizing correlated 
gapless topological states
and materials.
Here, low-energy excitations emerge from strong correlations and are subjected to space-group symmetry constraints, leading to correlated topological phases in theoretical models and materials that realize such phases.
  {\bf b,\,c,} The lattice structure of the square-net layer with sublattices $A$, $B$ on Wyckoff position $2c$. Also shown are the mirror ($M_z$) and screw ($C_x$ and $C_y$) nonsymmorphic symmetries. 
{\bf d,} $\mathcal{AA}$ stacking of the square-net layers into a 3D structure.
 }
 \label{fig:latt}
\end{figure}

These developments motivate the search for a general non-perturbative framework to treat 
the  interplay of correlation and topology
 and design both phases and materials of correlated gapless electronic topology.
Our hypothesis is that strong correlations cooperate with crystalline symmetry to produce such states.
Specifically, as Fig.\,\ref{fig:latt}(a) illustrates,
strong correlations give rise to emergent
excitations {\it at low energies.}
The space group symmetry constrains these excitations, leading to 
 emergent topological 
 phases while enforcing their gapless nature.
If validated, the proposition 
provides a materials design principle for correlated gapless electronic topology.
Still, the
 proposed cooperation is counter-intuitive, because strong correlations tend to cause localization
and 
gap out electronic excitations \cite{Morimoto2016,Wagner2021}. 

To test the proposed
materials design principle,
it is important to go beyond 
toy models
and explore correlated systems with general space group symmetries that 
may promote different types of nodal electronic states.
We illustrate our approach by narrowing
down the choice of 
strong-coupling models for our study here
as follows.
We focus on space group symmetries where electronic degeneracies 
may develop at partial fillings, regardless of orbital content.
These often happen in nonsymmorphic space groups.
We examine the case where interactions do not break crystalline or time-reversal symmetries, 
as a point of principle. 
Finally, to be definite, 
we formulate our materials design in the context of Kondo lattice systems,
as realized via the limit of strong
Coulomb repulsion in periodic Anderson models.
The Kondo effect leads to a ground state
 in which the local moments and conduction electrons
entangle and form a spin singlet, which supports composite fermions in the excitation spectrum 
at low energies -- within
the Kondo energy scale of the Fermi energy \cite{Louren2018,Ahamed2018,Jang2019}.
Generalizations to other correlated systems will be 
discussed.

We thus consider a Kondo lattice system that contains
mirror symmetry,
which is known to favor Weyl nodal lines in noninteracting systems \cite{Chiu2014,Bian2015,Bian2016,Chan2016}.
Our focus is on the nonsymmorphic and noncentrosymmetric square-net systems,
which host both mirror and screw nonsymmorphic symmetries [Fig.\,\ref{fig:latt}(b)(c)].
Especially, space group (SG) 129
has been a fertile setting for electronic degeneracies 
 of noninteracting
 systems
 \cite{Young2015,Schoop2016,Schoop2018,Muechler2020,Klemenz2020,Nica2015}
and is particularly advantageous in that the
degeneracies are orbital-independent \cite{Bradlyn2017}.
Allowing for inversion-symmetry breaking, we demonstrate the cooperation of strong correlations with space group symmetry in producing 
a novel phase -- a Weyl-Kondo nodal-line semimetal. 
Our analysis of correlation-driven topological semimetal phases gives rise to
a general procedure to design new materials
that realize such 
 phases. 
 We illustrate this materials design principle by identifying several new candidate Ce-based correlated-topological semimetals, and indicate how it can be used to identify many new materials.

The periodic Anderson model (PAM) is specified by the following Hamiltonian:
\begin{equation}
\label{eq:pam}
\mathcal{H} = \mathcal{H}_d + \mathcal{H}_{cd} + \mathcal{H}_c \, .
\end{equation}
The spin-$1/2$ $d$- and $c$-fermion operators describe the 
physical localized $f$ electrons 
and the light $spd$ conduction electrons that form the noninteracting bands,
respectively. Further details of the model and solution method are described in the Methods.

\begin{figure}[t]
\includegraphics[width=0.85\columnwidth]{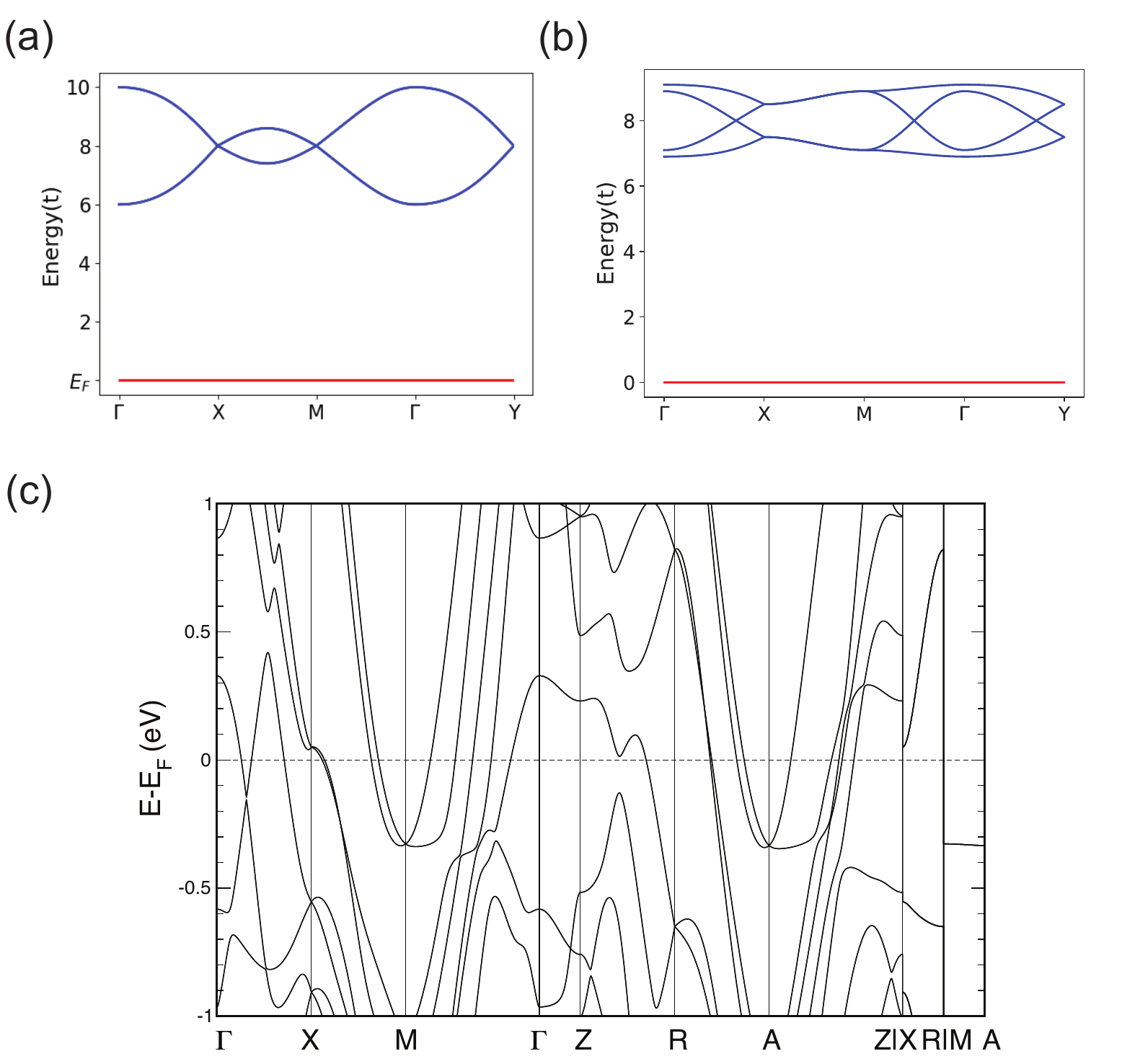}
\vskip 0.15in
\caption{
{\bf Electronic structure in the absence of Kondo effect.}
{\bf a,} Dispersion of a 2D square net with ($t_1, t_2, t^{SO}$) = ($1, 0, 0.6$)).  
The line at $E_F$ illustrates the localized electrons.
The four-fold degeneracies of the conduction bands occur at high symmetry points $X, Y, M$ away from the Fermi energy. 
{\bf b,} Dispersion of the 3D-stacked square net 
without either
 inversion symmetry breaking or SOC. 
The parameters are ($t_1, t_2, t_{z}^1, t_z^{2}, \Delta_z$) = ($1, 0.4, 0.3, 0.3, 0.8$).
{\bf c,} 
Bandstructure of the conduction $spd$-electrons of
CePt$_2$Si$_2$, 
 as determined by $f$-core DFT calculations. 
Symmetry-dictated Dirac nodes appear at $X$ and $M$ (and $R$ and $A$) 
away from the Fermi energy,
 as captured by the model dispersions of {\bf a} and {\bf b}.
}
\label{fig:bare}
\end{figure}

The 3D crystalline structure of SG 129 ($P4/nmm$)
 and its inversion-breaking counterpart, SG 31, are constructed by stacking 2D layers
in the $z$-direction. 
To set the stage for 
analyzing the interplay between space group symmetry and 
correlation effects, we start from this 2D system corresponding to the layer group $p4/nmm$.
The Hamiltonian of the conduction 
electrons, $\mathcal{H}_c$, is described in
a matrix form:
$\mathcal{H}_c=\sum_{{\bf k}} \Psi_{{\bf k}}^{\dagger} H_c({\bf k}) \Psi_{\bf k}$, 
where $\Psi_{\bf k}^{T} = (c_{{\bf k}\uparrow A}, c_{{\bf k}\downarrow A}, c_{{\bf k}\uparrow B}, c_{{\bf k}\downarrow B})$, 
with $A$ and $B$ being the two sublattices, ${\bf k}$ the wavevector, and $\sigma=\uparrow, \downarrow$ marking the 
spin quantum numbers.
As described in the Methods
(and illustrated in Fig.\,\ref{fig:stack_full}),
it contains 
tight-binding hopping terms
between the nearest ($t_1$) and next-nearest ($t_2$) neighbors,
an SOC ($t^{SO}$) term, and an inversion-symmetry-breaking ($\Delta$) term.
In the absence of the Kondo effect,
all the electrons prefer to occupy the 
low-lying
$d$-states to half filling, making the $d$-component to be a Mott insulator. 
Concomitantly, 
the conduction $c$-electron bands are completely empty, {\it i.e.} they lie above the Fermi energy. 
Figure\,\ref{fig:bare}(a) shows the bandstructure for the case 
of the inversion-symmetry-breaking potential $\Delta=0$,
{\it i.e.}
in the presence of
the full 
$p4/nmm$ symmetry.
The mirror and screw nonsymmorphic symmetries enforce additional crossings at high symmetry points 
$X$, $Y$ and $M$ (see Methods)~\cite{Young2015}. 
Since these Dirac points occur far above the Fermi energy, they leave the ground state topologically trivial.
We note that the separated $d$-electrons are deep levels far away (below) the Fermi energy; they are half-filled 
and form a Mott insulator due to the large onsite Coulomb repulsion.
Figure\,\ref{fig:2dkondo}(a) shows the 
dispersion
of the Kondo-driven, Fermi-energy bound, composite fermions for $\Delta=0$.
Because the composite fermion bands are subjected
 to the nonsymmorphic symmetry constraint, they 
 feature Dirac nodes at $X$, $Y$ and $M$.

We are now in position to
present our main results on the 3D SGs 129 and 31.
 For SG 129, the mirror and screw symmetries have the same effect as in the 2D case.
 This implies that  Kondo-driven,  dispersive, Dirac nodes robustly develop along the
 XR, MA and YS lines in the composite-fermion spectrum within the Kondo energy of the Fermi energy.
 Their dispersive nature implies that a Dirac nodal point can develop at the Fermi energy. 
 We illustrate this point through
 a concrete calculation 
 [Supplementary Information (SI),
 Figs.\,\ref{fig:aa}(a)(b)]. We note that, in the presence 
 of a magnetic field (or ferromagnetic order), Kondo-driven 
 Weyl nodes develop 
 [SI, Figs.\,\ref{fig:aa}(c)(d)].

\begin{figure}[t]
\includegraphics[width=0.95\columnwidth]{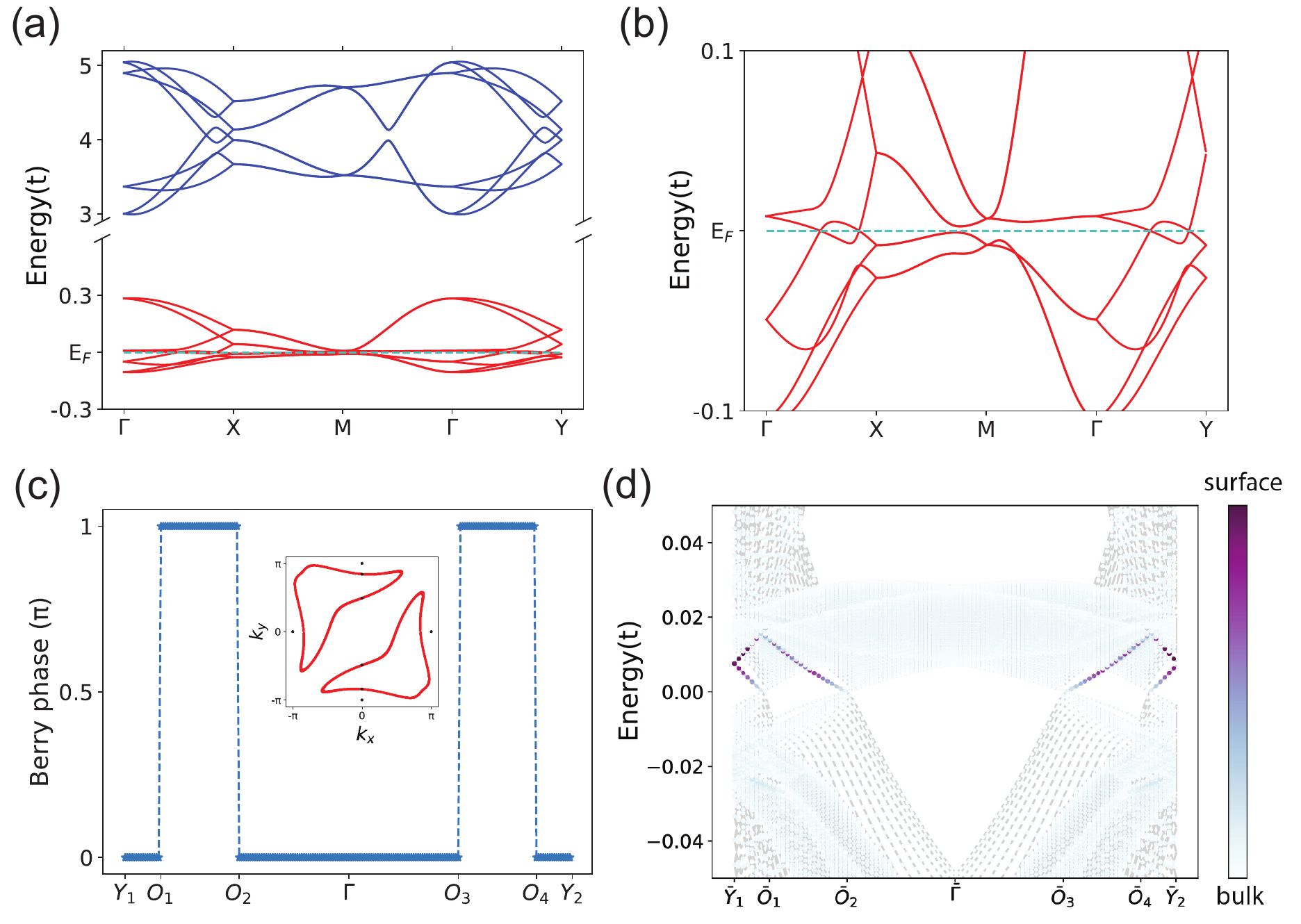}
\vskip 0.15 in
\caption{
{\bf 
Kondo-driven composite fermions
of 3D stacked square-nets with broken inversion symmetry.}
{\bf a,} The dispersion, including for the
composite fermions (the lower red part), 
within the $k_z=0$ plane. Here $V=3$, in unit of $t_1$ that is set to $1$. The other parameters are set to
$(t_2,\Delta,t_z^1,t_z^2, \Delta_z,m_{z},E_d,\mu)=(0.4, 0.4, 0.3, 0.3, 0.8,0.08,-2,-3.730)$. 
The parameters were chosen without any fine-tuning; the only requirement is that $r_1$ and $r_2$ are 
found to be nonzero. The saddle-point analysis yields
$(r_1, r_2, l_1, l_2)=(0.498, 0.374, -2.820, -2.294)$. 
{\bf b,} The zoomed-in dispersion of the 
Kondo-driven composite fermions.
{\bf c,} 
The Berry phase (in the $z$-direction) of the Kondo-driven Weyl nodal lines for fixed $k_x = 0$; $O_{1,2,3,4}$ mark
the points of intersection with the nodal lines [Fig.\,\ref{fig:stack_full}(c)].
Inset: the Kondo-driven Weyl nodal lines
 in the $k_z = 0$ plane.
{\bf d,} Kondo driven drumhead surface states. Shown here is the
momentum-resolved surface density of states for the spectrum close to Fermi energy on the $(001)$ surface along 
$Y_1(0, -\pi)$ to $Y_2(0, \pi)$.
A bar atop a momentum represents its projection onto the surface.
}
\label{fig:3d}
\end{figure}

 The case of SG 31 is more involved.
 To be definite, we consider $\mathcal{AA}$
  stacking of square-net layers as illustrated in Fig.\,\ref{fig:latt}(d),
 with the two stacking layers respectively hosting $s$ and $p$ orbitals \cite{Bian2015,Bian2016}.
 The corresponding periodic Anderson model is given in the Methods.
In the presence of the Kondo effect,
our results are shown
 in Fig.\,\ref{fig:3d}(a).
Nodes
  develop in the
dispersion of the Kondo-driven
 composite fermions, 
which are more clearly seen in 
Fig.\,\ref{fig:3d}(b)
as we zoom in to 
the immediate vicinity of the Fermi energy.
In our model, the Kondo-driven nodes
are
precisely at $E_F$.
These electronic degeneracies appear in the form of Weyl-nodal lines in the $k_z=0$ plane,
as shown in Fig.\,\ref{fig:3d}(c). All these reflect the 
 mirror nonsymmorphic $\{ M_z|\frac{1}{2} \frac{1}{2}0\}$ 
symmetry. 
Because
they are associated with the highly renormalized composite fermions,
the Weyl nodal-line excitations have strongly reduced velocities.
We stress that the nodal lines are robust against the SOC. 

The realization of the Weyl-Kondo nodal-line semimetal 
is particularly important, as it suggests
the emergence of drumhead surface states. We have 
computed the excitation spectrum on the (001) face in a slab with $40$ unit cells. 
As indicated in Fig.\,\ref{fig:3d}(d), we find  Kondo-driven
drumhead surface states that are bounded
 by the projections of the Weyl rings. Importantly, the dispersion of the
drumhead states also captures the Kondo energy scale and, hence, is strongly correlation-renormalized.
 
We now turn to the experimental signatures of the Kondo-driven nodal-line Weyl semimetals. 
The first category of the signatures bear similarities with
those of the Weyl-Kondo semimetals \cite{Lai2018,Grefe2020_2,Grefe2021,Dzsaber2017,Dzsaber2021,Dzsaber2019.x}. 
For example,
 because the velocity $v^*$ is highly reduced from the typical bare conduction-electron value $v$,
the specific heat is expected to have a quadratic temperature 
dependence with a large enhancement factor (see the SI):
\begin{equation}
c_v = \Gamma \,T^2 \, .
\label{eq:c-linear-in-T2}
\end{equation}
Here, the prefactor is
$\Gamma  = \frac{K}{2\pi}
\frac{9 \zeta (3)}{2} \left ( \frac{k_B}{\hbar v^*} \right )^2 k_B$,
where $K$ is the length of the nodal line and the zeta function $\zeta(3) \approx 1.202$.
It  is enhanced from the typical noninteracting value by $(v/v^*)^2$, which is
of the order $(W/k_BT_K)^2$, where 
$W$ is
the bare conduction-electron
bandwidth.
This enhancement factor is huge--about $10^4-10^6$ for moderately to strongly renormalized heavy-fermion semimetals.
As another example,
the enhanced Berry curvature near the Fermi energy can be probed through the spontaneous Hall effect
\cite{Dzsaber2021}.

The second category of experimental signatures are distinct for the Kondo-driven nodal-line phase.
As one example, the nodal lines lead to a non-trivial Berry phase of any closed loop 
perpendicular to the mirror plane. 
The resulting characteristic signatures in quantum oscillations for such a strongly correlated setting are
 considerably 
more challenging to probe than their weakly correlated 
counterparts \cite{Li2018_osc,Yang2018,Kwan2020}. However, 
in materials with moderate mass enhancement
(as we will identify below), 
these experiments are expected to be feasible.
As another example, the drumhead surface states will lead to distinct spectroscopic signatures. 
They can, as in weakly correlated systems\cite{Stuart2022},
 in principle be probed by quasiparticle-interference measurements via scanning tunneling microscopy (STM);
 here, the quasiparticle-interference pattern develops structure
  at wavevectors whose magnitude is bounded from above 
 by the location of the nodal lines in the bulk, as well as
 at energies that are bounded above 
 by the bandwidth of the drumhead states \cite{Biderang2018}.
 Finally, developments of recent years \cite{Kirchner2020_rmp}
 make angle-resolved photoemission spectroscopy (ARPES) as a promising
 probe of the dispersive bulk and surface $f$-electron states we have discussed.

We turn next to demonstrating how our approach guides the search for new correlation-driven topological materials.
Because  the composite fermions must be located near the Fermi energy,
we can expect the Kondo-driven semimetal phases to host topological nodal excitations near the Fermi
energy for generic fillings. Consequently,
 the cooperation of symmetry and Kondo correlation
 is adequate to realize candidate 
Kondo materials; this
general procedure is outlined in Fig.\,\ref{fig:latt}(a).
In passing, we note on one general point.
Designing strongly correlated topological materials is inherently difficult; in
 the presence of strong correlations, 
{\it ab initio} calculations of electronic states represent a challenge. Here, we bypass this difficulty
by using symmetry.
Our results on the model Kondo lattice Hamiltonians imply that space group symmetry and Kondo correlation
cooperate in driving correlated topological semimetals. Hence, we can design new materials for 
correlation-driven topological semimetals based purely on crystalline symmetry and strong correlations, 
without resorting to
 {\it ab initio} results for the correlated electronic structure. 

\begin{figure}[t]
\includegraphics[width=1.03\columnwidth]{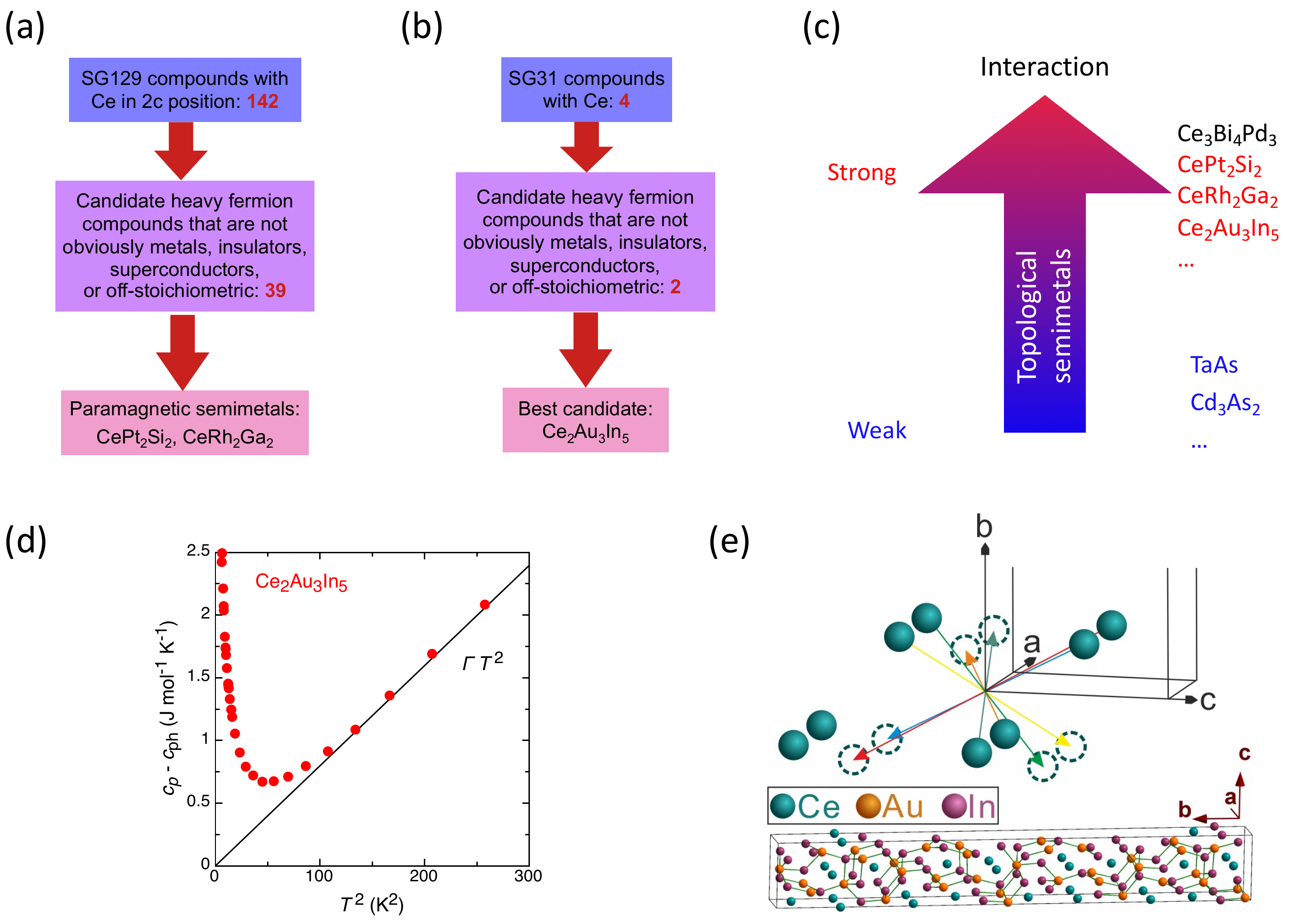}
\vskip 0.15 in
\caption{
{\bf Design of new materials for correlation-driven topological semimetal phases
and the first synthesized material.}
The design procedure as applied to SG 129 with Ce-ions on the Wyckoff position 2c ({\bf a}) and to SG 31 ({\bf b}).
The 39 compounds of the middle box in {\bf a} and the 4 compounds of {\bf b}
 are given in the Supplementary Information (Fig.\,\ref{fig:materials-details}).
{\bf c,} Summary of the newly identified candidate materials (red) 
for correlation-driven topological semimetals based on 
the search in
{\bf a} and {\bf b} and their placement along the correlation axis.
{\bf d,} Electronic specific heat, $c_p - c_{ph}$, as a function of $T^2$. 
In the paramagnetic regime 
sufficiently above the low-temperature upturn (possibly a tail of a lower-lying magnetic phase transition),
a $\Gamma T^2$ form is observed.
{\bf e.} Crystal structure of Ce$_2$Au$_3$In$_5$ and illustration of its inversion-symmetry breaking.
}
\label{fig:materials}
\end{figure}

We first consider SG 129, and focus on the case of the
Ce ions on site 2c. The procedure is outlined 
in Fig.\,\ref{fig:materials}(a) and further described in the
SI. It leads to two new materials,
CePt$_2$Si$_2$ and CeRh$_2$Ga$_2$,
which we propose
to realize the Kondo-driven semimetal phase.
Both remain paramagnetic 
 down to the lowest temperature
of somewhat below 2 K \cite{Anand2017,Gignoux1986}
that has been experimentally measured,
and both have strong correlations as inferred from their moderately 
enhanced specific heat (see Supplementary Information).
Moreover, both show semimetalic behavior.
Below their respective Kondo temperatures,
their resistivity as a function of temperature 
[Fig.\,\ref{fig:rss}(a)(b)] behaves similarly as the well-established heavy-fermion semimetals
Ce$_3$Bi$_4$Pd$_3$ and CeNiSn \cite{Dzsaber2017,Nakamoto1995},
respectively [Fig.\,\ref{fig:rss}(c)(d)].
For CePt$_2$Si$_2$, the $spd$ electronic structure, determined by $f$-core DFT calculations
(see Methods), are displayed in Fig.\,\ref{fig:bare}(c). 
As can be seen, the symmetry of SG 129 
dictates the existence of Dirac nodes at the $X$, $M$, $R$ and $A$ points, which are located 
away from the Fermi energy.
The results for CeRh$_2$Ga$_2$ are similar (see the SI).
The DFT-calculated electronic structure is expected to apply well above the Kondo temperature. Below their respective 
Kondo temperatures, the approach we have advanced here suggests that they will realize Kondo-driven topological 
semimetals. The precise nature of the renormalized band structure requires the construction of tight-binding representations 
of the DFT-derived band structure as well as {\it ab initio} input of the Kondo couplings.
We outline how this construction can be done in the 
SI, and reserve correlated {\it ab initio} studies for a future work.
Still, our solution of the model Kondo lattice Hamiltonian with SG 129 and the resulting conclusion 
about the cooperation between
the symmetry and Kondo correlation imply that both materials qualify as candidates for the 
correlation-driven topological semimetals advanced here.

Studies of these materials set the stage to search for additional strongly correlated semimetals.
In the case of a nonzero inversion-symmetry-breaking term $\Delta$,
corresponding to SG 31, 
a similar search procedure for the Ce-based case
is outlined in Fig.\,\ref{fig:materials}(b) and in the 
SI. It leads to the identification of a new material,
Ce$_2$Au$_3$In$_5$, as
a candidate heavy-fermion material that is known
not to order down to the lowest measured temperature ($2$ K) \cite{Galadzhun1999}.
Its inversion-symmetry breaking is illustrated in Fig.\,\ref{fig:materials}(e).
We propose it as a candidate material
 to realize the Weyl-Kondo nodal-line semimetal advanced here.
 Thus, considerations of these two specific space groups
 lead to the identification of three new candidate Kondo-driven topological
 semimetals [Fig.\,\ref{fig:materials}(c)]. This result already considerably expands the material base for such strongly correlated 
 topological semimetals beyond the known case of Ce$_3$Bi$_4$Pd$_3$ \cite{Dzsaber2017,Dzsaber2021,Dzsaber2019.x}.
 
 To demonstrate the predictive power of our materials design principle, we have synthesized 
 single-crystalline Ce$_2$Au$_3$In$_5$ and its La-counterpart,
 La$_2$Au$_3$In$_5$. The electronic specific heat of
  Ce$_2$Au$_3$In$_5$ is shown in Fig.\,\ref{fig:materials}(d)
  (see the SI for how the phonon contribution is determined).
The low-temperature upturn indicates the influence of a (presumably magnetic) phase transition 
below the lowest measured temperature ($2$\,K). In the paramagnetic regime free of this influence,
a $\Gamma T^2$ form is observed. Fitting $\Gamma$ in terms of 
Eqs.\,(\ref{eq:c-linear-in-T2},\ref{c_expression_Gamma_v*}) yields 
a nodal velocity $v^* = 2994$\,m/s. This represents a close-to $3$-orders of magnitude renormalization from the typical
velocity of noninteracting electrons, which is consistent with the association with Kondo-driven composite fermions.

Our approach for materials design can be readily generalized.
For example, our procedure applies to other Wyckoff positions (e.g., Ce ions at positions 2a or 2b instead of 2c), 
 other lanthanide elements (Pr, Sm, Eu, Yb, in addition to Ce) and actinide elements (e.g., U) and, finally,  
 a large number of other space group symmetries (in particular, those among the 155 nonsymmorphic 
 space groups). These factors make it likely that hundreds  of
 candidate materials can be realized for strongly correlated topological semimetals phases.

We now underscore some general lessons
drawn from our results. Our models have an even number of electrons per unit cell. 
 The typical localizing tendency of strong correlations could have turned the systems into Kondo insulators \cite{AeppliFisk,Dzero2010,Nikolic2014}.
Instead, the strong correlations cooperate with the space group symmetries [Fig.\,\ref{fig:latt}(a)]:
the Kondo effect generates composite fermions; the space group symmetry
constraints on the composite fermions prevent the Kondo gap from developing
 and lead instead to the Weyl nodal excitations.
 
Going beyond Kondo systems, emergent
 low-energy excitations can already develop for intermediate correlations,
when the correlation strength is comparable to the noninteracting bandwidth.
Here too, they will be  subjected
to space group symmetry constraints.  
 To illustrate the point, consider a multi-orbital Hubbard model containing both Hubbard and Hund's 
 interactions and with the orbitals having unequal kinetic energies \cite{Yu17.1,Komijani17.1}.
When a correlated metal develops near an
orbital-selective Mott phase,
the emergent low-energy excitations
take the form of a narrow band that is bound to the immediate vicinity of the Fermi energy,
as indeed seen experimentally \cite{Huang2022}.
For such models in suitable crystalline settings, 
the space group symmetries are expected to constrain the low-energy excitations  and produce 
gapless topological phases. Thus, our work motivates
parallel (and systematic) studies for correlated gapless electronic topology
in such multi-orbital models and the associated transition-metal compounds.
We can expect our approach to provide a means for identifying
hitherto-unknown phases and materials of correlated electronic topology
 in a variety of settings across a wide correlation spectrum. 
Finally, correlation physics that involves nontrivial electronic 
topology is a general problem that pertains to a growing list of materials. 
As recent studies \cite{Song2021}
in the moir\'{e} bands of twisted bilayer graphene illustrate, the kind of topological heavy fermions
we advance here may well serve as a platform to elucidate the enigmatic physics of these emerging materials.

In conclusion, we have advanced a materials
design principle for strong correlations to cooperate
with space group symmetry and drive correlated topological solids.
Strong correlations lead to low-energy emergent excitations, which are subject to the constraints of the 
space group symmetry for topological phases.
We have tested this approach in Kondo lattice models whose space group symmetries 
may promote different kinds of electronic degeneracies.
In nonsymmorphic and noncentrosymmetric square-net systems, we have theoretically validated
the approach by demonstrating novel Weyl-Kondo nodal-line semimetals in both two and three dimensions. 
The approach has allowed us to propose a general procedure to identify new materials for 
correlation-driven topological semimetals, to apply it and 
identify a number of such 
materials in several representative space groups, and to suggest that many others can be designed in this way.
We have synthesized single-crystalline Ce$_2$Au$_3$In$_5$ and provided new experimental results to support the 
theoretical prediction.
Our findings illustrate the potential of the proposed materials design principle to
guide the search for new correlated topological metals in a broad range of strongly correlated quantum materials.

\clearpage

\vspace{0.3cm}
\noindent{\bf Acknowledgments}\\
Work at Rice has been supported by the 
Air Force Office of Scientific Research under Grant No.\ FA9550-21-1-0356 (C.S. and Q.S.),
the National Science Foundation under Grant No.\ DMR-2220603 (L.C.)
and the Robert A. Welch Foundation Grant No.\ C-1411 (H.H.).
The majority of the computational calculations have been 
performed on the Shared University Grid at Rice funded by NSF under Grant EIA-0216467, 
a partnership between Rice University, Sun Microsystems, and Sigma Solutions, 
Inc., the Big-Data Private-Cloud Research Cyberinfrastructure MRI-award funded by NSF under Grant No. CNS-1338099,
and the Extreme Science and Engineering Discovery Environment (XSEDE) by NSF under Grant No.\  DMR170109.
M.G.V. acknowledges the support from the Spanish Ministry of Science and Innovation Grant No.\ 
PID2019-109905GB-C21 
and the Deutsche Forschungsgemeinschaft (DFG, German Research Foundation) 
GA 3314/1-1 -- FOR 5249 (QUAST).
Work at Los Alamos was carried out under the auspices of the U.S. Department of Energy (DOE) National 
Nuclear Security Administration under Contract No.\ 89233218CNA000001, and was supported by LANL LDRD Program.
J.C. acknowledges the support of the National Science Foundation under Grant No.\ DMR-1942447 
and the support of the Flatiron Institute, a division of the Simons Foundation.
Work in Vienna was supported by
the Austrian Science Fund (projects No.\  29279-N27
and I 5868-N - FOR 5249 - QUAST).
Four of us (S.G., S. P., J. C. and Q.S.) acknowledge the hospitality of the Aspen Center for Physics,
which is supported by NSF grant No. PHY-1607611.

\vspace{0.2cm}
\noindent{\bf Author contributions}
\\
Q.S., J.C. and S. P. conceived the research. L.C., C.S., H.H., S.E.G., J.C. and Q.S. carried out theoretical model studies.
A.P. and S.P. identified candidate materials for the proposed correlated topological semimetals. 
L.F. and X.Y. synthesized the material and G.E. performed the specific heat measurements.
M.G.V. performed DFT calculations.
L.C., C.S., H.H., J.C. and Q.S. wrote the manuscript, with inputs from all authors.

\vspace{0.2cm}
\noindent{\bf Competing 
 interests}\\
The authors declare no competing 
 interests.
 
 \vspace{0.2cm}
 \noindent{\bf Additional information}\\
Correspondence and requests for materials should be addressed to 
Q.S. (qmsi@rice.edu), J.C. (jennifer.cano@stonybrook.edu) and S.P. (paschen@ifp.tuwien.ac.at)

\clearpage
\clearpage

\clearpage

\clearpage

\clearpage

\noindent{\bf\large Methods}\\
\\
{\bf The periodic Anderson model, solution method and symmetry analyses}~~
The PAM is specified by Eq.\,\ref{eq:pam}.
The model contains two species of spin-$1/2$
electrons. The Hamiltonian $\mathcal{H}_d$,
containing spin-$1/2$ $d$-operators, describes 
the physical localized $f$ electrons:
\be
\mathcal{H}_d = E_d \sum_{i,\sigma} d^{\dagger}_{i\sigma} d_{i\sigma} + U\sum_i n^{d}_{i\uparrow} n^{d}_{i\downarrow} \, ,
\ee
with the energy level $E_d$ and onsite Coulomb repulsion $U$.
Whereas $\mathcal{H}_c$,
involving $c$-operators,
characterizes the
$spd$ conduction electrons that form the noninteracting bands.
The hybridization term, which describes the Kondo coupling between the conduction and $f$- electrons,
is given by
\be
\mathcal{H}_{cd} = V\sum_{i\sigma} \left( d_{i\sigma}^{\dagger}c_{i\sigma} + h.c. \right) \, ,
\ee
where $V$ represents 
the hybridization between the two species of electrons.
The following features of the symmetry group 129 and model justify its application to many $f$-electron materials 
in this space group.
The symmetry representations of this space group are such that at each high-symmetry point, 
the degeneracy of the bands is independent of the symmetry of the orbital (assuming spin-orbit coupling). 
Thus, our model applies regardless of the nature of the orbitals near the Fermi energy.
The dispersion of the Kondo-driven composite fermions
stays near the Fermi energy, within the Kondo energy.
 Thus, while we specialize in the quarter-filled case, corresponding to one electron per site,
 changing the filling will only shift the composite-fermion bands up to the Kondo energy.
 The saddle-point analysis, involving
self-consistent equations that are exact in a large-$N$ limit (where $N$ is the spin degeneracy),
 is described in  the Supplementary Information (SI).
The Kondo effect generates composite fermions that must be 
located near the Fermi energy. 
While we will describe symmetry constraints in terms of the dispersion of the composite 
fermions, the same conclusion is reached through a more general analysis that is carried out 
in terms of the eigenvectors 
of the full single-particle Green's function for the interacting Hamiltonian \cite{Hu2021} (see SI).

The PAM 
in 2D involves
$\mathcal{H}_c=\sum_{{\bf k}} \Psi_{{\bf k}}^{\dagger} H_c({\bf k}) \Psi_{\bf k}$, 
where $\Psi_{\bf k}^{T} = (c_{{\bf k}\uparrow A}, c_{{\bf k}\downarrow A}, c_{{\bf k}\uparrow B}, c_{{\bf k}\downarrow B})$, and
\be
\label{eq:cond}
\begin{aligned}
H_{c}({\bf k}) = & t_1 \cos \frac{k_x}{2} \cos \frac{k_y}{2} \tau_{x}  +t_2(\cos k_x + \cos k_y)
+ t^{SO} ( \sin k_{x}
\sigma_y  -\sin k_y 
\sigma_x) \tau_z
 \\ & 
+ \Delta \sin(\frac{k_x-k_y}{2}) \tau_x \sigma_z,
\end{aligned}
\ee
where ${\bf \tau} = (\tau_x, \tau_y, \tau_z)$'s and ${\bf \sigma} = (\sigma_x, \sigma_y, \sigma_z)$'s 
are Pauli matrices acting on the (A,B) sublattice and spin spaces respectively.
A more explicit form of the Hamiltonian can be found in the SI.
For definiteness, we consider the lattice model with tight-binding hopping terms 
between the nearest ($t_1$) and next-nearest ($t_2$)
sites [as illustrated in the SI, Fig.\,\ref{fig:stack_full}(a)],
an SOC ($t^{SO}$) term, and an inversion-symmetry-breaking ($\Delta$) term.
The first three terms 
contain all the symmetry-allowed components up to the next-nearest neighbor,
describing an effective $s$-orbital model 
on Wyckcoff position $2c$ 
with both time reversal 
symmetry ($\mathcal{T}$) and inversion symmetry ($\mathcal{I}$). 
The presence of both $\mathcal{T}$ and $\mathcal{I}$ ensure each band is doubly degenerate with states 
$| g \rangle$ and $\mathcal{T} \mathcal{I} | g\rangle$. In addition, the 
symmetries include
the glide mirror symmetry $\{ M_z| \frac{1}{2} \frac{1}{2}\}$ (Fig.\,\ref{fig:latt}(b)(c))) and two screw symmetries
 $\{ C_{2x}| \frac{1}{2} 0\}$ and $\{ C_{2y}| 0 \frac{1}{2} \}$ 
(where $\{ C_{ni}| {\bf t}\}$ = n-fold rotation along axis $i$ + fractional translation ${\bf t}$ as depicted in Fig.\,\ref{fig:latt}(c)).
The bands are generically two-fold degenerate. However,
a four-fold degeneracy at $X$ arises as a result of either the mirror nonsymmorphic $\{ M_z| \frac{1}{2} \frac{1}{2}\}$ symmetry or the screw nonsymmorphic $\{ C_{2x}| \frac{1}{2} 0\}$ symmetry. The same degeneracy 
appears at $Y$,
due to either the mirror nonsymmorphic $\{ M_z| \frac{1}{2} \frac{1}{2}\}$ symmetry
or the screw nonsymmorphic
$\{ C_{2y}| 0 \frac{1}{2} \}$ 
symmetry. 
Finally, a four-fold degeneracy also develops at $M$, from
either the $\{ C_{2x}| \frac{1}{2} 0\}$ or the
$\{ C_{2y}| 0 \frac{1}{2} \}$ symmetry.
 The fourth term has the form of $\Delta \sin(\frac{k_x-k_y}{2}) \tau_x \sigma_z$.
It breaks the inversion symmetry of $p4/nmm$,
while preserving its mirror symmetry \cite{Young2015,Wu2019}.
 The quarter-filling case we examine corresponds to
 one electron per site.
We consider the realistic regime of parameters corresponding to the 
limit of strong coupling, $U \rightarrow \infty$, and with the $d$-electron level 
being sufficiently deep compared 
to the energy of the conduction-electron bands ({\it i.e.}, for
sufficiently
negative $E_d$), in which the system is in the Kondo limit. 
The saddle-point results are compatible with those derived from a variety of other methods.
For example, the Kondo effect is found to develop only when the hybridization strength goes above a threshold value.
This reflects the fact that the bare conduction-electron bath has a gap near the Fermi energy. The gap
makes the Kondo coupling to be irrelevant (as opposed to marginally relevant) in the renormalization-group
sense; as a result, a nonzero threshold value of the Kondo coupling or hybridization is needed for the 
Kondo effect to develop.

We now specify the PAM in 3D. Here,
 $\mathcal{H}_d$ and $\mathcal{H}_{cd}$ 
 are the same (except that the site 
 summation is in the 3D lattice).
For the conduction
electrons,
the Hamiltonian now takes the form:
$\mathcal{H}_c^{3D} 
=  \sum_{\bf k} \Phi_{\bf k}^{\dagger} H_c^{3D}({\bf k}) \Phi_{\bf k}$, where $\Phi_{\bf k}^{T}=(c_{{\bf k}\uparrow A}^s, 
c_{{\bf k}\downarrow A}^s, c_{{\bf k}\uparrow B}^s, c_{{\bf k}\downarrow B}^s, 
c_{{\bf k}\uparrow A}^p, c_{{\bf k}\downarrow A}^p, c_{{\bf k}\uparrow B}^p, c_{{\bf k}\downarrow B}^p)$.
In this equation,
\begin{equation}
\label{eq:pam3D}
\begin{aligned}
H^{3D}_c({\bf k})  
= H_c(k_x, k_y) \otimes \rho_z + \Delta_z \rho_z + t_z^1 \sin(\frac{k_z}{2}) \rho_y
+ t_{z}^2\cos(k_z) \rho_z +m_z \tau_x \otimes \sigma_x \otimes \rho_y 
\, ,
\end{aligned}
\end{equation}
with $\rho_{x,y,z}$ being the Pauli matrices in the space of the two stacking layers
(see the 
SI
for a more explicit form of the Hamiltonian).
Here, the third and the fourth terms represent the nearest- and next-nearest-neighbor hoppings
along the $z$-direction 
[as illustrated in the SI, Fig.\,\ref{fig:stack_full}(a)]
and the fifth term is the SOC along the $z$-direction. 
For simplicity, we have chosen an $s$- and $p$-orbital stacking and
set the $s$ and $p$ orbitals 
to couple with the $d$-electrons with the same hybridization strength.
However,
the nature of the resulting topological semimetal phase is determined by the underlying 
space-group symmetry and is generic to the given SG.
Further details of our analysis is described in the 
SI.
Again we focus on the quarter-filling case. 

\noindent
{\bf Crystal growth and specific heat of Ce$_2$Au$_3$In$_5$ and La$_2$Au$_3$In$_5$}~~
Single crystals of Ce$_2$Au$_3$In$_5$ were grown from In flux using alumina crucibles.
Elements of high purity ($>$$99.9\%$) were taken for the synthesis. Single crystals
were separated from the melt by centrifuging. Attempts to prepare by this way
single crystals of La$_2$Au$_3$In$_5$ failed. A single phase polycrystalline 
La$_2$Au$_3$In$_5$
sample was prepared by melting pure elements in the stoichiometric ratio in a
high frequency furnace and annealing the as cast sample at 
$ 650^\circ$C for three days.
All samples were analysed by x-ray powder diffraction and energy dispersive 
x-ray spectroscopy (EDX) for phase purity and composition prior to the physical
property measurements. The single crystallinity of the Ce$_2$Au$_3$In$_5$ samples was
checked by Laue x-ray measurements.
The specific heat $c_p$ was measured using the Quantum Design heat capacity
option for the PPMS. 
Figure\,\ref{fig:CeLa235}(a) shows the $c_p$ of Ce$_2$Au$_3$In$_5$ and La$_2$Au$_3$In$_5$
as a function of $T^3$. The phonon contribution to the specific heat $c_{ph}$, 
assumed as usual to be the same for both compounds, was estimated
from the $c_p$ of La$_2$Au$_3$In$_5$ by subtracting its electronic contribution, 
determined as follows:
The $c_p/T$ of La$_2$Au$_3$In$_5$ at low temperatures
($<$$ 5$\,K) is well described by a
 conventional Debye-Sommerfeld model, $c_p/T = \gamma + \beta T^2$. It yielded
a finite electronic specific heat coefficient $\gamma=(16.2$$ \pm$$ 0.6)$\,mJ/mol K$^2$ and a 
$\beta$ value that corresponds to the Debye temperature $(232$$ \pm $$5)$\,K. 
The electronic specific heat of Ce$_2$Au$_3$In$_5$ was obtained by
subtracting this $c_{ph}$ from its total $c_p$.

\noindent
{\bf $f$-core DFT calculations}~~
The electronic structure calculations were performed with the $f$ states removed from the pseudo-potential to model the band structure.
The SOC was taken into account. Further details can be found in the 
SI.

\noindent
{\bf Data availability}~~
The data that support the findings of this study are available from the corresponding author
upon reasonable request.

\clearpage

\setcounter{figure}{0}
\setcounter{equation}{0}
\makeatletter
\renewcommand{\thefigure}{S\@arabic\c@figure}
\renewcommand{\theequation}{S\arabic{equation}}

\noindent{\bf\Large Supplementary Information}\\
\\
\noindent {\bf Kondo-driven topological semimetals in 2D}\\
The Kondo effect leads to composite fermions near the Fermi energy. 
For the inversion-symmetric case  ($\Delta=0$), the renormalized 
dispersion as shown in Fig.\,\ref{fig:2dkondo}(a) was already discussed in the main text.
For the case with broken inversion symmetry (i.e., with nonzero $\Delta$),
we show  the renormalized dispersion  in  Fig.\,\ref{fig:2dkondo}(b). 
The four-fold degenerate states at $X$ and $Y$ are split into two pairs of two-fold Weyl nodes above 
and below the Fermi energy. 
The
different branches of the Weyl nodes cross each other at the Fermi energy,
giving rise to Fermi-energy-bound 
Weyl-nodal lines, as shown in 
Fig.\,\ref{fig:2dkondo}(c). The Weyl-nodal lines are protected by the mirror nonsymmorphic symmetry
\cite{Young2015,Wu2019}. 
We note that, here too, the Kondo-driven nodal lines have been identified in the presence of SOC.

\noindent{\bf Kondo-driven topological semimetals for SG 129}\\
Utilizing the 2D model as the building block, we consider how the Kondo-driven 
topological nodes  develop in the SG 129 case.
This can already be done with a direct stacking along the $z$-direction. 
As shown in Fig.\,\ref{fig:aa}(a),
the symmetry enforced Dirac points at high symmetry points in 2D is 
extended along $z$-direction and becomes Dirac nodal lines along the high symmetry hinges;
the degree of their dispersion is related to the strength of the $z$-direction hoppings.
Thus, there could only be Dirac points that cross the Fermi energy [Fig.\,\ref{fig:aa}(b)].
In the presence of a ferromagnetic order or Zeeman coupling from an external magnetic field, the Dirac nodes are turned into Weyl nodes.
In addition, pairs of Weyl nodes can in principle appear along $\Gamma-X/Y/Z$, which are 
protected by rotational symmetries according to magnetic group analysis. 
This is illustrated 
in Fig.~\ref{fig:aa}(c)(d), which show a pair of 
Weyl nodes along $\Gamma-Z$.

\noindent {\bf Additional data for the Kondo-driven nodal line in 3D}\\
In  Fig.\,\ref{fig:3d} of the main text, we showed the renormalized electronic structure for the 3D square
net along high-symmetry directions within the $k_z = 0$ plane.  
Here, we plot the bands along high-symmetry directions in the entire 3D Brillouin zone. The renormalized band structure is shown 
in Fig.~\ref{fig:stack_full}(a) along the path connecting high symmetry points as can be seen through Fig.~\ref{fig:stack_full}(b). 
Except for the Weyl line in the $k_z=0$ plane, the heavy bands are fully gapped across the 3D Brillouin zone.

\noindent {\bf Hamiltonians in matrix form}\\
Consider first the 2D case.
The Hamiltonian of the conduction $c$-fermion, $\mathcal{H}_c$, is described in
a matrix form:
$\mathcal{H}_c=\sum_{{\bf k}} \Psi_{{\bf k}}^{\dagger} H_c({\bf k}) \Psi_{\bf k}$, 
where $\Psi_{\bf k}^{T} = (c_{{\bf k}\uparrow A}, c_{{\bf k}\downarrow A}, c_{{\bf k}\uparrow B}, c_{{\bf k}\downarrow B})$, 
with $A$ and $B$ being the two sublattices, ${\bf k}$ the wavevector, and $\sigma=\uparrow, \downarrow$ marking the 
spin quantum numbers.
\begin{equation*}
H_{c}(\bf{k}) =
    \mx
    f_2({\bf k}) & f_{SO}({\bf k}) & f_1({\bf k}) + f_{\Delta}({\bf k}) & 0 \\
    f_{SO}({\bf k}) & f_2({\bf k}) & 0 & f_1({\bf k}) - f_{\Delta}({\bf k}) \\
    f_1({\bf k}) + f_{\Delta}({\bf k}) & 0 & f_2({\bf k}) & -f_{SO}({\bf k})\\
    0 & f_1({\bf k}) - f_{\Delta}({\bf k}) & -f_{SO}({\bf k}) & f_2({\bf k})
    \ex
\end{equation*}
with
\begin{equation*}
\begin{aligned}
    f_1({\bf k}) &= t_1 \cos \frac{k_x}{2} \cos \frac{k_y}{2}, \\
    f_2({\bf k}) &=t_2(\cos k_x + \cos k_y), \\
    f_{SO}({\bf k}) &= t^{SO}(-\sin k_y - i\sin k_x), \\
    f_{\Delta}({\bf k}) &= \Delta \sin(\frac{k_x-k_y}{2}),    
\end{aligned}
\end{equation*}
where $t_1/t_2$ are the nearest and next-nearest neighbor hopping parameters, 
$t^{SO}$ is the symmetry-allowed SOC and $\Delta$ breaks the inversion symmetry.

Consider next the Hamiltonian for the 3D model.
For the conduction $c$-fermions, 
the Hamiltonian now takes the form:
$\mathcal{H}_c^{3D} 
=  \sum_{\bf k} \Phi_{\bf k}^{\dagger} H_c^{3D}({\bf k}) \Phi_{\bf k}$, where $\Phi_{\bf k}^{T}=(c_{{\bf k}\uparrow A}^s, 
c_{{\bf k}\downarrow A}^s, c_{{\bf k}\uparrow B}^s, c_{{\bf k}\downarrow B}^s, 
c_{{\bf k}\uparrow A}^p, c_{{\bf k}\downarrow A}^p, c_{{\bf k}\uparrow B}^p, c_{{\bf k}\downarrow B}^p)$. And
\begin{equation*}
    H_{c}^{3D}(\bf{k}) = \mx
    H_c(k_x, k_y) + H_{z}(k_z) & \rho(k_z) \\
    \rho^{\dagger}(k_z) & - H_c(k_x, k_y) - H_{z}(k_z)
    \ex,
\end{equation*}
where 
\begin{equation*}
    H_z(k_z) = (\Delta_z + t_z^2\cos k_z) \mathds{1}_{4\times4},
\end{equation*}
and
\begin{equation*}
    \rho(k_z) = \mx
    -it_z^1\sin{\frac{k_z}{2}} & & &-im_z \\
     &-it_z^1\sin{\frac{k_z}{2}} &-im_z  & \\
     &-im_z  &-it_z^1\sin{\frac{k_z}{2}} & \\
     -im_z  & & &-it_z^1\sin{\frac{k_z}{2}}
    \ex.
\end{equation*}
Here, $\Delta_z$ is the chemical potential difference between two layers, $t_z^1$/$t_z^2$ represent the nearest- and next-nearest-neighbor hoppings along the $z$-direction and $m_z$ is the SOC along $z$-direction.

\noindent {\bf Saddle point equations}\\
We describe the  auxiliary-boson method in this section.
In the 2D square-net model, we first fix the electron filling to be one per site 
\be
n_d + n_c = 1 \, ,
\label{eq:fill2d}
\ee
with
\ba
n_d & = \frac{1}{N_{\mathrm{u}}} \sum_{i,\sigma} d^{\dagger}_{i\sigma} d_{i\sigma} \, , \\
n_c & = \frac{1}{N_{\mathrm{u}}} \sum_{i,\sigma} c^{\dagger}_{i\sigma} c_{i\sigma} \, ,
\ea
where $N_{\mathrm{u}}$ counts the total number of sites.

In the infinite-$U$ limit, the localized $d$-electrons 
can be represented as $d^{\dagger}_{i\sigma} = b_i f^{\dagger}_{i\sigma}$ with local constraint
\be
b_{i}^{\dagger}b_{i} + \sum_{\sigma} f_{i\sigma}^{\dagger} f_{i\sigma} = 1 \, .
\label{eq:fill}
\ee
 This constraint rules out the double occupancy entirely. At the saddle point, 
  the auxiliary
  bosons 
  condense,
  with $b_i \to \langle b_i\rangle = r$. We further introduce a uniform Lagrangian multiplier $l$ to enforce the local constraint in Eq.\,\ref{eq:fill}, which gives rise to the saddle-point Hamiltonian:
 \be
 \mathcal{H}_s = \mathcal{H} + l \left( 
 \frac{1}{N_{\mathrm{u}}} 
 \sum_{i, \sigma} f_{i\sigma}^{\dagger} f_{i\sigma} + r^2 -1\right) - \mu
 \, 
 \frac{1}{N_{\mathrm{u}}} 
 \sum_{i,\sigma} n^{c}_{i\sigma},
 \ee
 where $\mu$ is the chemical potential. The saddle point equations are
\begin{align}
\sum_{\sigma} \left( V\langle 
c_{\sigma}^{\dagger}
 f_{\sigma}\rangle+h.c.\right) - 2rl&=0 \, , \\
 \sum_{\sigma} \langle f_{\sigma}^{\dagger} f_{\sigma}\rangle - r^2 &=1 \, .
\end{align}

Together  with the filling constraint Eq.\,\ref{eq:fill2d}, we can self-consistently determine the variational parameters $(\mu, r, l)$. 

We apply the same auxiliary-boson
method to the 3D stacked square net with a
saddle-point Hamiltonian
\begin{equation}
\begin{aligned}
H &= H_c^{3D} + r_1V 
\frac{1}{N_{u}}
\sum_{i,\sigma} (s_{i,\sigma}^{\dagger} f_{i,\sigma}^1 +h.c.) + E_d{\frac{1}{N_{u}}}
\sum_{{i,}\sigma} f_{{i,}\sigma}^{1\dagger}  f_{{i,}\sigma}^{1} \\
& + r_2V {\frac{1}{N_{u}}} \sum_{{i,}\sigma} (p_{{i,}\sigma}^{\dagger} f_{{i,}\sigma}^2 +h.c.) 
+ E_d {\frac{1}{N_{u}}} \sum_{{i,}\sigma} f_{{i,}
\sigma}^{2\dagger}  f_{{i,}\sigma}^{2} \\
& + l_1{\frac{1}{N_{u}}} \left( \sum_{{i,}\sigma} f_{{i,}\sigma}^{1\dagger} f_{{i,}\sigma}^{1}+ r_1^2-1\right) + l_2
{\frac{1}{N_{u}}} \left( \sum_{{i,}\sigma} f_{{i,}\sigma}^{2\dagger} f_{\sigma}^{2} + r_2^2-1\right) \\
&- \mu{\frac{1}{N_{u}}}  \sum_{{i,}\sigma} \left( s^{\dagger}_{{i,}\sigma}s_{{i,}\sigma} + p^{\dagger}_{{i,}\sigma} p_{{i,}\sigma}\right),
\end{aligned}
\end{equation}
where $f_{\sigma}^1$ and $f_{\sigma}^2$,  $r_1$ and $r_2$, $l_1$ and $l_2$ 
represent the 
auxiliary 
fermions, condensed 
auxiliary bosons and Lagrangian multipliers for each layer.

The saddle point equations are
\begin{align}
\frac{1}{N_{u}} \sum_{\sigma} \left( V\langle s_{\sigma}^{\dagger} f_{\sigma}^1\rangle+h.c.\right) - 2r_1l_1&=0 \, , \\
\frac{1}{N_{u}} \sum_{\sigma} \langle f_{\sigma}^{1\dagger} f_{\sigma}^1\rangle - r_1^2 &=1, \\
\frac{1}{N_{u}} \sum_{\sigma} \left( V\langle p_{\sigma}^{\dagger} f_{\sigma}^2\rangle+h.c.\right) - 2r_2l_2&=0 \, , \\
\frac{1}{N_{u}} \sum_{\sigma} \langle f_{\sigma}^{2\dagger} f_{\sigma}^2\rangle - r_2^2 &=1 \, .
\end{align}
The filling is  still fixed to be one per site. 

\noindent {\bf Symmetry analysis of the interacting Green's function}\\
We briefly note on the symmetry analysis of the interacting Green's function \cite{Hu2021}.
This utilizes the 
eigenvectors and eigenfunctions of the single-particle Green's function of an interacting multi-band system.
The symmetry analysis as applied on the eigenvectors proceeds in a similar way as for the Bloch functions 
of noninteracting bands. The degeneracies of the eigenvectors imply the degeneracies of the
peaks in the energy dispersion of the corresponding spectral function.
Further details can be found in Ref.\,\citenum{Hu2021}.

\begin{figure}[t]
\includegraphics[width=0.88\columnwidth]{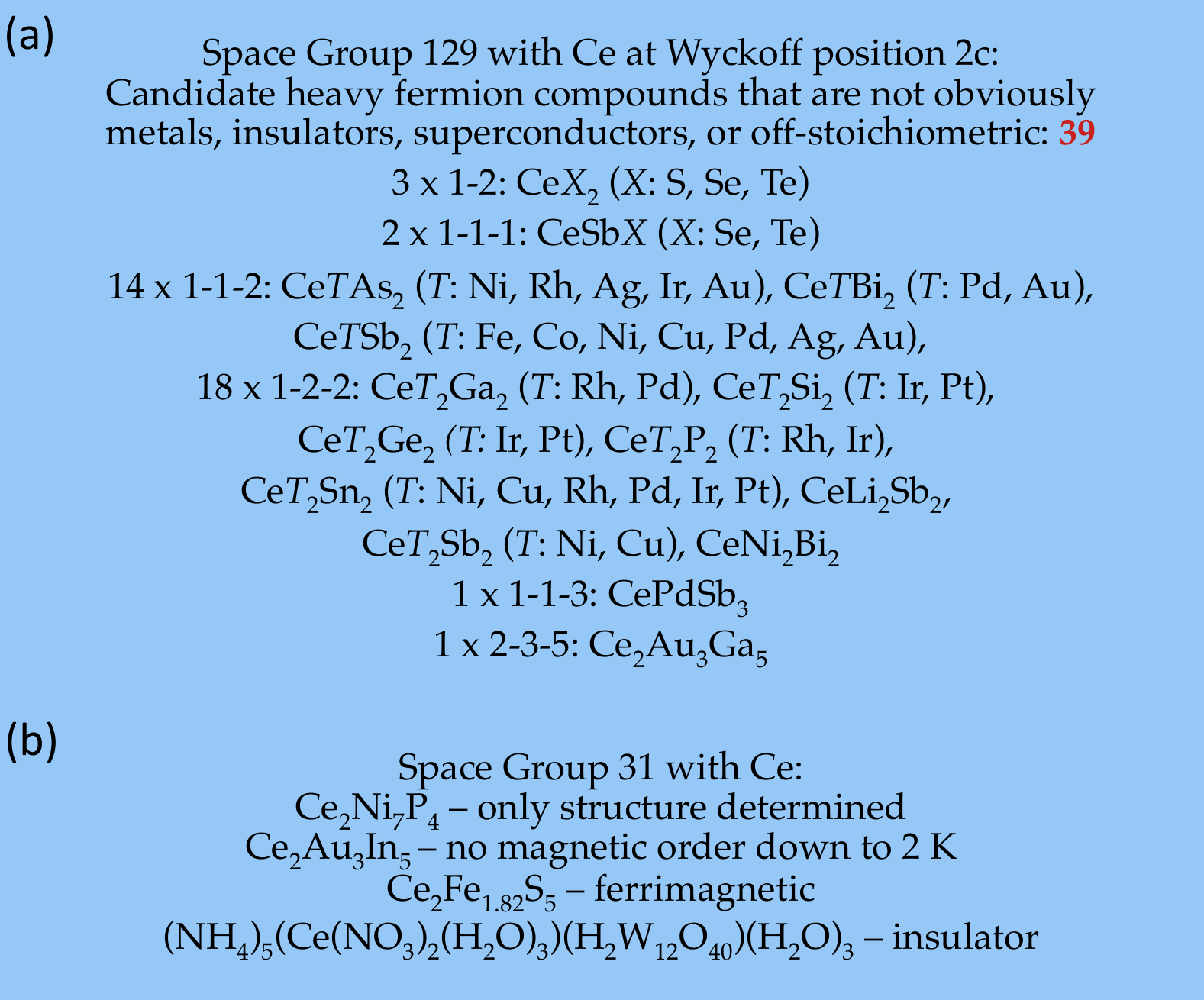}
\vskip 0.15 in
\caption{
{\bf Further details in the search for correlation-driven topological semimetals.}
Shown here are the materials identified during the intermediate step of the search procedure as outlined in 
Fig.\,\ref{fig:materials}(a) (main text) for SG 129 with Ce-ions on the Wyckoff position 2c ({\bf a})
and in Fig.\,\ref{fig:materials}(b) (main text) for SG 31 with Ce atoms ({\bf b}).
}
\label{fig:materials-details}
\end{figure}

\noindent {\bf Search for new materials to realize the proposed topological semimetal phases 
driven by strong correlations}\\
The search for candidate materials started with the retrieval of Ce
compounds crystallizing in the space group P4/nmm (SG 129) from the Inorganic
Crystal Structure Database (ICSD). Multiple entries were unified. Only materials
containing Ce atoms in the unique 2c Wyckoff position were taken into account.
This yields 142 chemical compounds as specified in Fig.\,\ref{fig:materials}(a).
In the second step,
the following entries were removed: Insulators (oxides, oxohalogenides,
oxosulfides), off-stoichiometric phases and solid solutions, 
obviously good metals (based on valence considerations) and 
superconductors and
related phases.
The remaining 39 pure
stoichiometric phases were predominantly ternary cerium compounds of pnictogens,
chalcogens, and the silicon group elements [Fig.\,\ref{fig:materials-details}(a)].
The further selection has to be done
by literature search and inspection of the measured physical properties (where
available), as illustrated in the next subsection.
Two examples resulting from such a search are identified as strongly correlated
paramagnetic semimetals, CePt$_2$Si$_2$ and CeRh$_2$Ga$_2$, and are proposed 
as materials for the correlation-driven topological semimetal phases advanced here.
A similar procedure applies to the case of SG 31 with Ce atoms. The 4 materials 
of this case are shown in Fig.\,\ref{fig:materials-details}(b).

We note in passing that magnetic materials are also of interest, even though they are not the 
focus of the present work. For example, in the case of SG 129,
CeSbTe is a well-known 
 material that hosts 
$f$-electrons \cite{Schoop2018}. However, it exhibits an antiferromagnetic order at $T_N=2.7$ K, which breaks 
time-reversal symmetry. Still, the amount of entropy that has been measured in such a temperature range suggests that
the system will be strongly correlated if it is tuned to
a paramagnetic state. 
Thus, tuning these systems into their paramagnetic ground  state 
is suggested as
a means of realizing the 
Kondo-driven topological semimetal we have advanced here.

\noindent {\bf CeRh$_2$Ga$_2$ and CePt$_2$Si$_2$ as strongly correlated semimetals}\\
The two candidate materials we have identified, CeRh$_2$Ga$_2$ and CePt$_2$Si$_2$, 
show properties 
that are characteristic 
of strongly correlated semimetals. Their strong correlations are 
signaled by the large enhancement 
of the specific heat from the typical values of weakly correlated metals.
The electronic specific-heat 
coefficient $C/T$ reaches
$130 $ and $80$ mJ/mol K$^2$, respectively \cite{Anand2017,Gignoux1986}
({\it i.e.} about 100 times the effective mass of weakly correlated bulk metals).
Their semimetal nature is captured by the resistivity as a function of temperature, respectively 
shown in Fig.\,\ref{fig:rss}(a)(b) for the two compounds \cite{Anand2017,Bhattacharjee1989}, 
which behaves similarly 
as Fig.\,\ref{fig:rss}(c)(d), respectively, for the well-established heavy-fermion semimetals
Ce$_3$Bi$_4$Pd$_3$ and CeNiSn \cite{Dzsaber2017,Nakamoto1995}.
In all cases, the temperature dependence of the resistivity is neither activated, as a fully-gapped system would be,
nor showing a giant $T^2$-dependent component at low temperatures with a positive temperature slope, 
as a typical
heavy-fermion metal would be. This feature is accompanied by the relatively large magnitude of the resistivity.

\noindent {\bf Further details and additional results of the DFT calculations}\\
The electronic structure calculations were performed using the
Vienna ab initio simulation package (VASP) \cite{Kresse1993,Kresse1996}
 with the Perdew, Burke and Ernz-erhof (PBE) generalized gradient approximation for the exchange correlation potential \cite{Perdew1996}.
The $f$ states were removed from the pseudo-potential to model the band structure. 
The interaction between the ion cores and valence electrons was described by
the projector augmented-wave method \cite{Kresse1999}.
The Hamiltonian contained the scalar relativistic corrections, and the SOC was taken into account by the second variation 
method \cite{Hobbs2000}.

\begin{figure}[t]
\centering
\includegraphics[width=0.98\columnwidth]{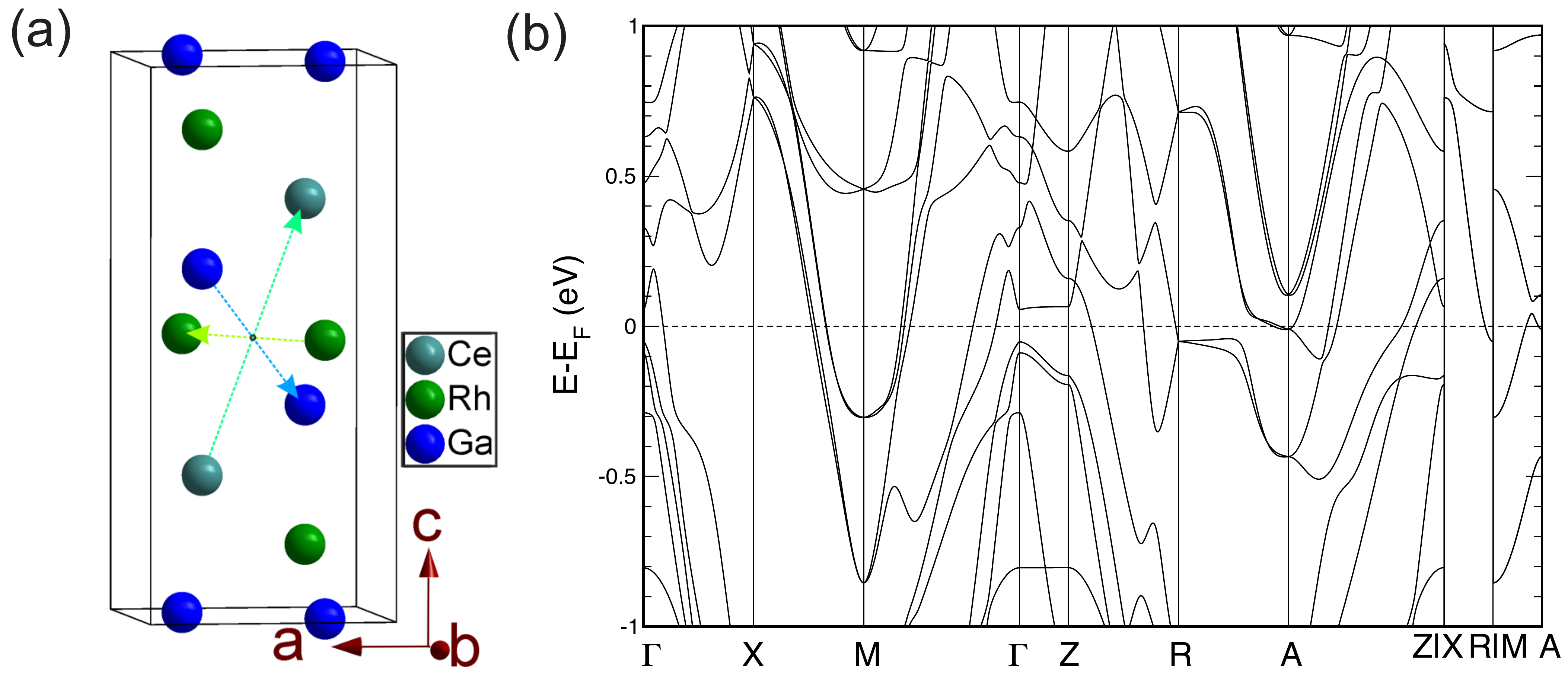}
\vskip 0.15 in
\caption{{\bf DFT results for $\mathrm{CeRh_2Ga_2}$.} {\bf a}, the crystal structure of $\mathrm{CeRh_2Ga_2}$;
{\bf b}, the DFT} results derived from ignoring the site-displacive disorder of the Rh and Ga atoms.
\label{fig:dft_crg}
\end{figure}

The calculation for 
CeRh$_2$Ga$_2$ is in principle more difficult, given
that the Rh- and Ga-site displacive disorder
in the existing materials has been reported~\cite{Nesterenko2017}. 
To get a qualitative understanding, we have carried out $f$-core DFT calculations
but ignoring the disorder, with the Rh and Ga atoms fully assigned to
the 2c and 2a sites, respectively. The results are shown in Fig.~\ref{fig:dft_crg}.

\begin{figure}[t]
\centering
\includegraphics[width=0.98\columnwidth]{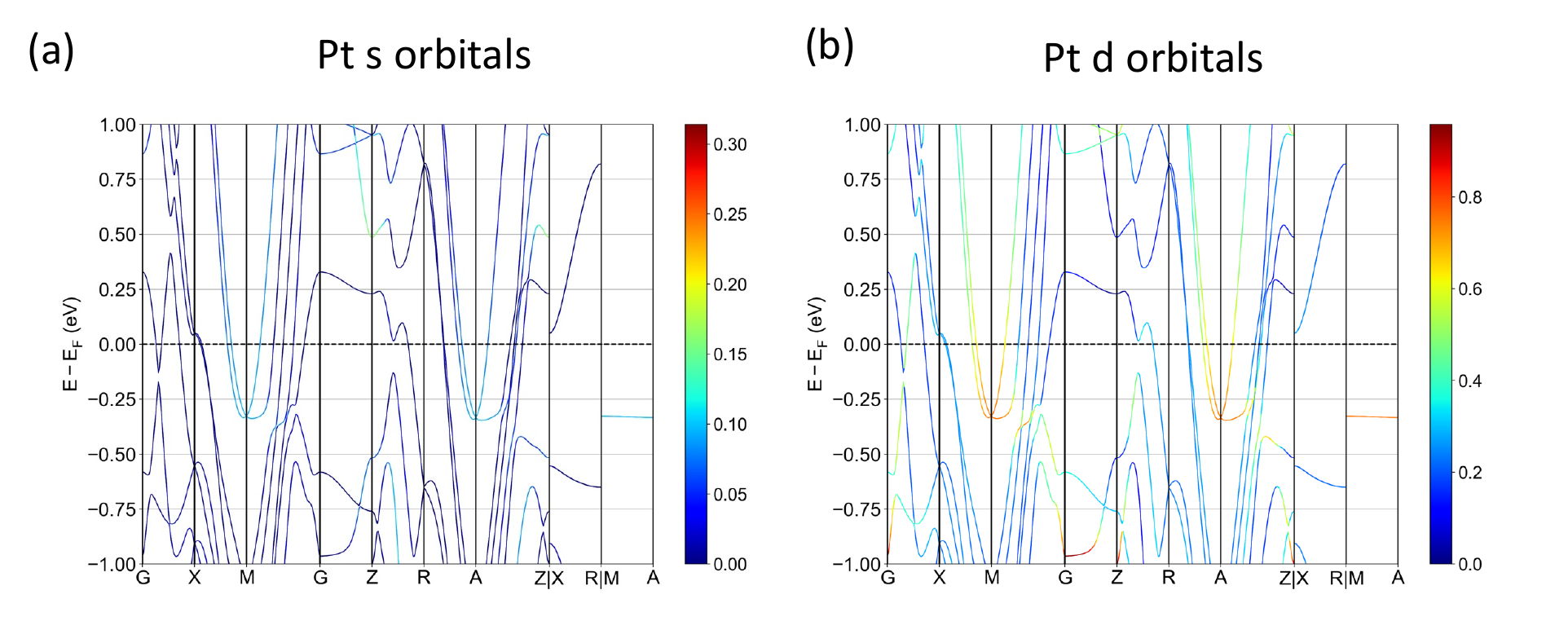}
\vskip 0.15 in
\caption{
{\bf DFT results for the atomic and orbital contents of the $spd$ conduction electrons in
$\mathrm{\bf CePt_2Si_2}$.}
The dominating contributions are found to come from {\bf a}, the Pt $s$-orbitals and {\bf b},
the Pt $d$-orbitals. 
The bands have been plotted using 
PyProcar \cite{Uthpala2020}.
}
\label{fig:dft_orb}
\end{figure}

A realistic periodic Anderson lattice model can be constructed as follows.
One can start from a
 tight-binding representation for the conduction electrons based on the understanding of the dominant orbitals
 that form the $spd$ conduction-electron bands; to illustrate the latter, we have analysed
 the 
atomic and orbital contents in the case of CePt$_2$Si$_2$, finding the dominating contributions to come 
from the Pt $d$- and $s$-orbitals (Fig.\,\ref{fig:dft_orb}). One also needs to determine the
 lowest Kramers doublet for the $f$-electrons, which is an interacting problem;
in practice, this usually requires input from spectroscopic measurements. 
Thereafter, one can construct the hybridization matrix.


\noindent {\bf Specific heat}\\
To calculate the low-$T$ 
specific heat, 
we focus on the linear-dispersion regime where we can approximate the dispersion 
$\epsilon_{\bf k} = \epsilon_{\bf k_{\perp}} = \hbar v^* k_{\perp}$.
Here, $ {\bf k} = (k_{\parallel}, {\bf k_{\perp}})$ represents the decomposition of the wave vector into
the components parallel and perpendicular to the nodal line;
 ${\bf k_{\perp}}$ itself has two components. 

The specific heat  becomes
\begin{equation}
\begin{aligned}
c_v &= \frac{\partial}{\partial T} \int_{\infty}^{+\infty} \frac{d\epsilon}{(2\pi)^3} \epsilon \rho(\epsilon)f(\epsilon) \\
& = \frac{\partial}{\partial T} \int_{-\infty}^{+\infty} \frac{d \epsilon}{(2\pi)^3} \epsilon \frac{2K\pi \epsilon}{(\hbar v^*)^2} \frac{1}{e^{\epsilon/k_BT}+1} \\
& = \frac{K}{(2\pi)^3} \frac{\partial }{\partial T} \int_{0}^{2\pi}d\theta \int_{0}^{k_K} dk_{\perp}  \frac{2\pi (\hbar v^*)^3k_{\perp}^2}{(\hbar v)^2} \frac{1}{e^{\hbar v^* k_{\perp}/k_BT}+1} \\
& = \frac{K}{2\pi}\frac{9\zeta(3)}{2}k_B \left( \frac{k_BT}{\hbar v^*}\right)^2 \, ,
\end{aligned}
\end{equation}
where $K$ is the length of the nodal line and $\zeta(3)\approx 1.202$. So the specific heat per unit volume is
\begin{eqnarray}
&c_v  = \Gamma \, T^2 \, , \\
&\Gamma = \frac{K}{2\pi} \frac{9\zeta(3)}{2} \frac{(k_B)^3}{(\hbar v^*)^2} \,.
\label{c_expression_Gamma_v*}
\end{eqnarray}

\clearpage

\begin{figure}[t]
\includegraphics[width=0.95\columnwidth]{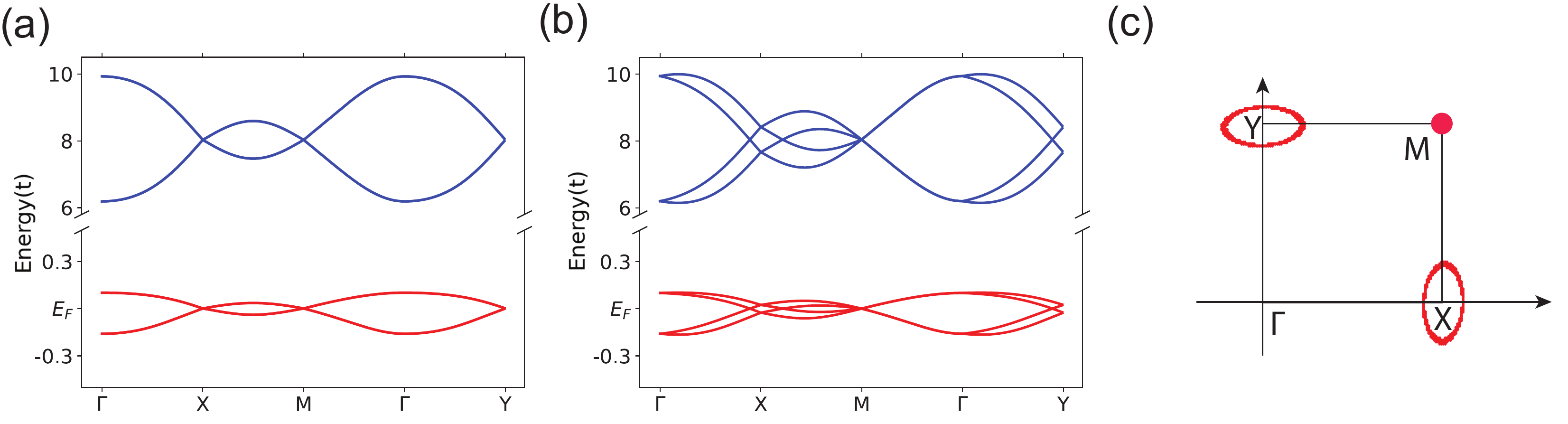}
\vskip 0.15 in
\caption{ 
{\bf 
Kondo-driven composite fermions of a 2D square net.}
{\bf a,}  The dispersion, including for the 
composite fermions (the lower red part), at $\Delta=0$.
The Kondo-driven Dirac nodes are pinned to $E_F$ at  the
high-symmetry points $X, Y, M$.
Here, the choice of parameters and the saddle-point analysis parallel
those of Fig.\,\ref{fig:3d}(a). The input parameters are $(V,E_d, t_1, t_2, t^{SO}, \mu)
= (6.8,-6, 1, 0, 0.6, -7.529)$. The saddle-point analysis yields $(r,l)=(0.285, -6.500)$.
{\bf b,} The dispersion of the
 Kondo-driven Weyl nodal-line semimetal.
{\bf c,} The Kondo-driven Weyl nodal lines  (the red lines)
encircle high symmetry points $X$ and $Y$. 
Here $\Delta=0.4$, and the other parameters are $(V,E_d, t_1, t_2, t^{SO}, \mu, r, l)$ = $(6.8,-6, 1, 0, 0.6, -7.564, 0.286, -6.500)$.
}
\label{fig:2dkondo}
\end{figure}
\clearpage

\begin{figure}[t]
\centering
\includegraphics[width=0.98\columnwidth]{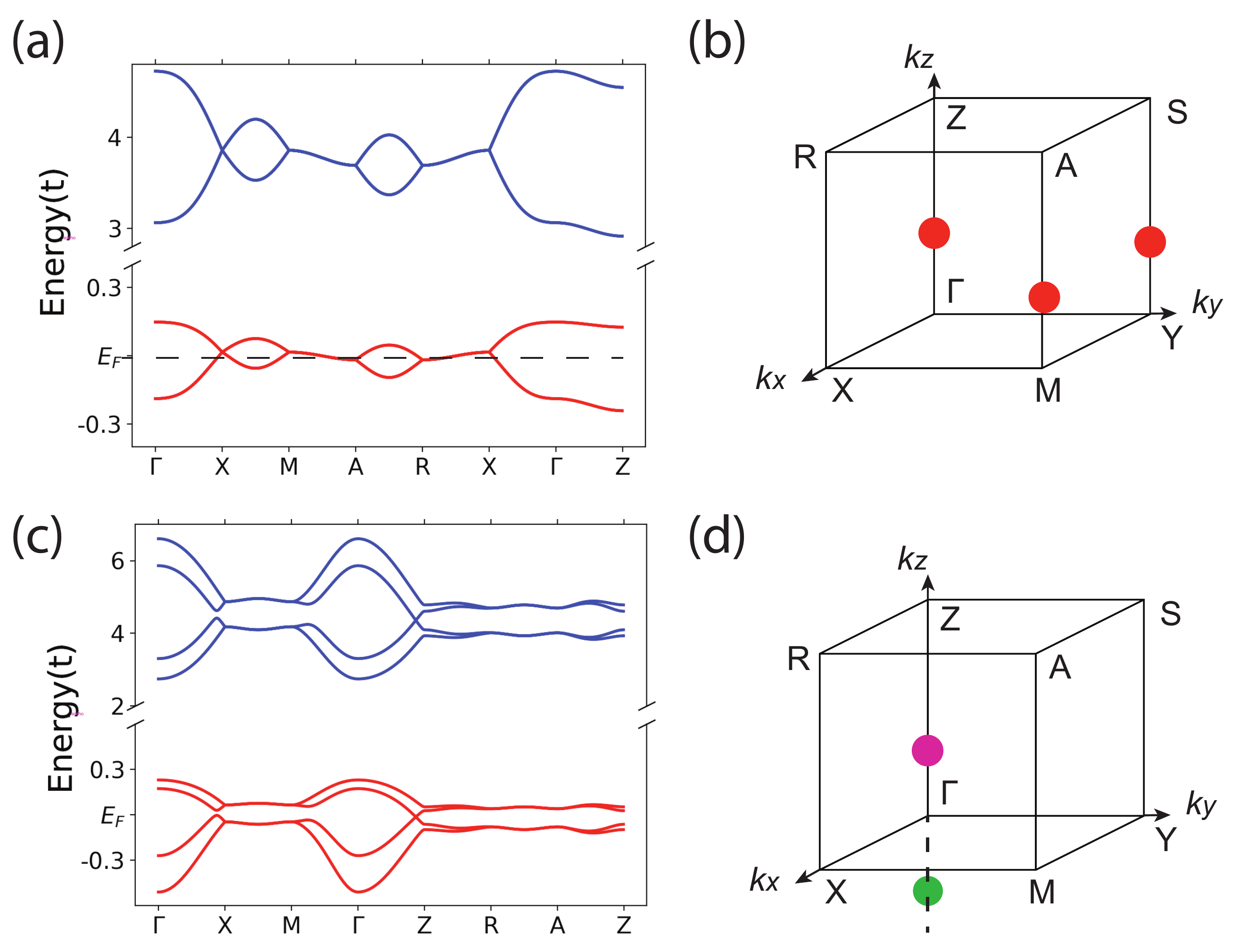}
\vskip 0.15 in
\caption{
{\bf Kondo driven composite fermions in the 3D model for SG 129.}
{\bf a},
The dispersion, includes a symmetry enforced dispersive
 Dirac nodal line along $X-R$, $Y-S$ and $M-A$; and 
 {\bf b},
the points on the Dirac nodal line that cross the Fermi energy. 
The parameters for
both {\bf a} and {\bf b} are  $(t_1, t_2, t^{SO}, t_{1z}, t_{3z}, m_z, \Delta, V, E_d, \mu, r, l) 
= (1, 0, 0.4, 0.1, 0, 0, 0, 3.6, -2.6, -3.70, 0.44, -3.13)$;
{\bf c}, in the presence of a magnetic field,
symmetry allows the existence of a pair of Weyl points 
along $\Gamma-Z$ as shown in
 {\bf d}. 
The parameters are $(t_1, t_2, t^{SO}, t_{1z}, t_{3z}, m_z, \Delta, V, E_d, \mu, r, l) 
= (1, 0, 0.4, 0.1, 0.8, 0.3, 0, 3.6, -2.6, -3.82, 0.46, -3.53)$.}
\label{fig:aa}
\end{figure}
\clearpage

\begin{figure}[t]
\centering
\includegraphics[width=0.98\columnwidth]{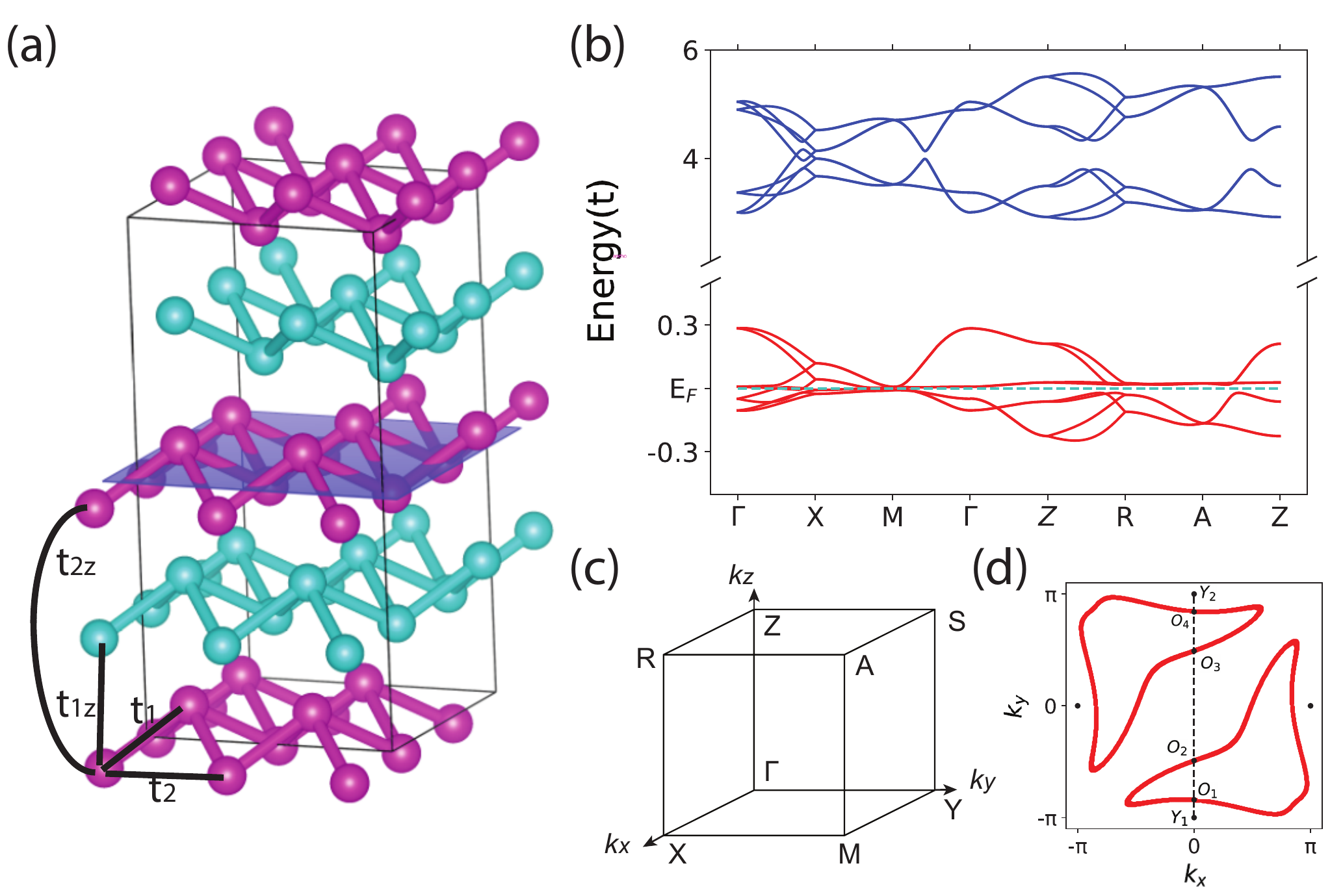}
\vskip 0.15 in
\caption{{\bf 
Kondo-driven composite fermions
 in the 
 3D model for SG 31
 over an extended region of the Brillouin zone.}
{\bf a},
Illustration of the hopping parameters in the 3D-stacked model;
{\bf b,} 
The dispersion across both the $k_z=0$ and $k_z=\pi$ planes 
with the same parameters as shown in Fig.\,\ref{fig:3d}. The
composite fermions
(lower red part)
are fully gapped when $k_z\neq0$. 
{\bf c},
The 3D Brillouin zone with high symmetry points $\Gamma=(0,0,0)$, $X=(\pi,0,0)$, 
$Y=(0,\pi,0)$, $M=(\pi,\pi,0)$, $Z=(0,0,\pi)$, $R=(\pi,0,\pi)$, $S=(0,\pi,\pi)$ and $A=(\pi,\pi,\pi)$.
{\bf d,}
 The Kondo-driven Weyl nodal lines, which intersect with the dashed $k_x=0$ line from $Y_1=(0,-\pi,0) $ to $Y_2=(0,\pi,0)$ 
  at $O_1$, $O_2$, $O_3$ and $O_4$.
}
\label{fig:stack_full}
\end{figure}
\clearpage

\clearpage

\begin{figure}[t]
\centering
\includegraphics[width=0.98\columnwidth]{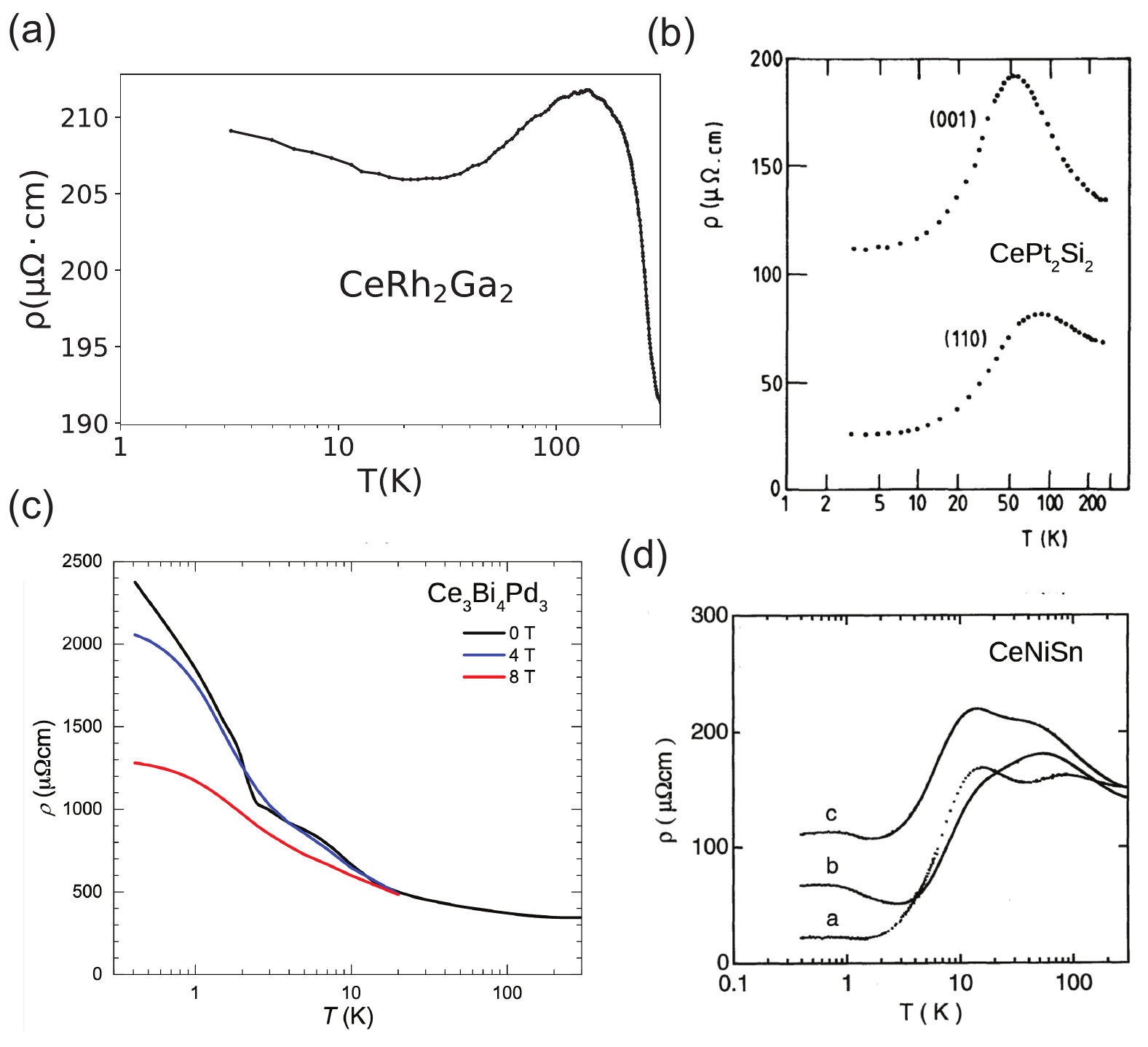}
\vskip 0.15 in
\caption{{\bf Semimetal transport of Ce-based square-net materials.}
Resistivity ($\rho$)
{\it vs.}
 temperature ($T$) of {\bf a,}
$\mathrm{CeRh_2Ga_2}$ 
{(replotted from Ref.\,\citenum{Anand2017})},
{\bf b,} $\mathrm{CePt_2Si_2}$~\cite{Bhattacharjee1989}, 
{\bf c,} $\mathrm{Ce_3Bi_4Pd_3}$~\cite{Dzsaber2017},
 and  {\bf d,} $\mathrm{CeNiSn}$~\cite{Nakamoto1995}.
 {All plots are shown in semi-log form.}
}
\label{fig:rss}
\end{figure}
\clearpage

\begin{figure}[t]
\centering
\includegraphics[width=0.98\columnwidth]{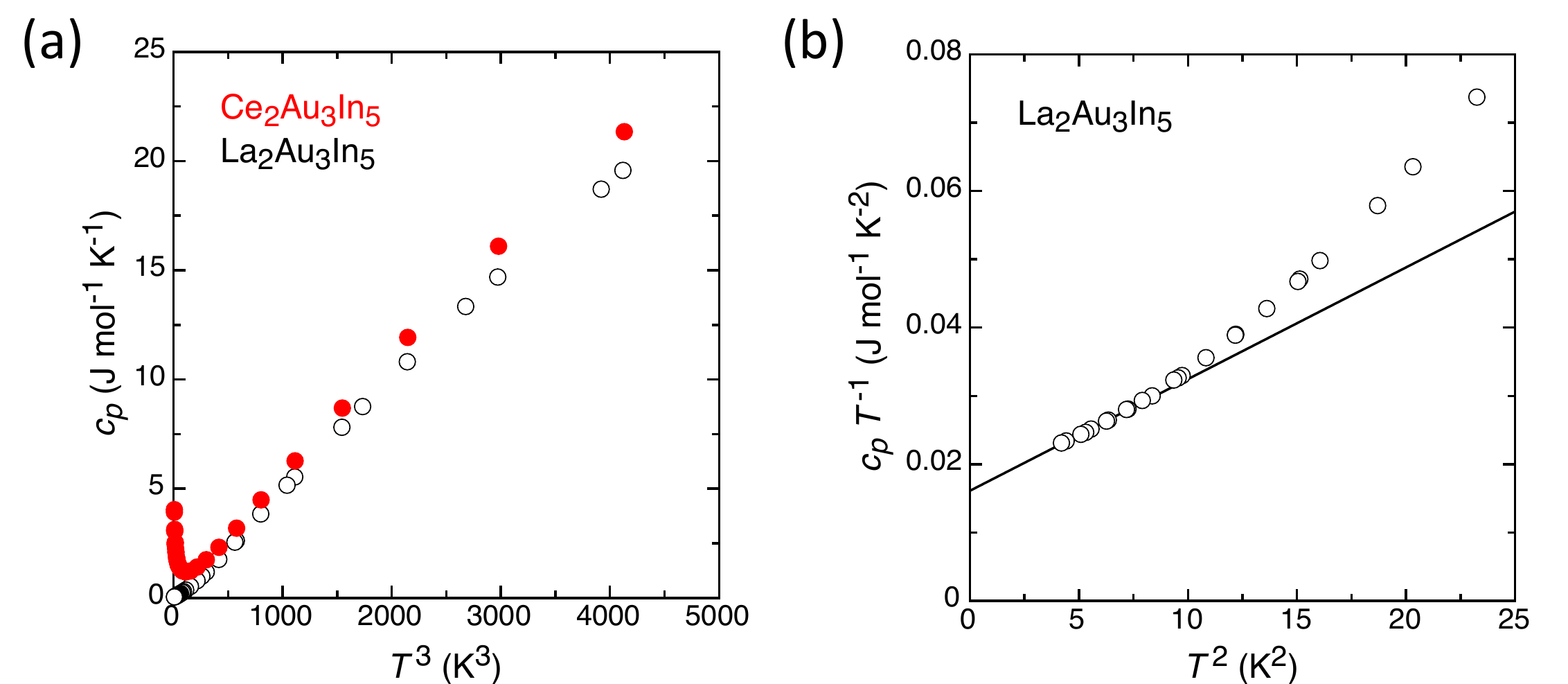}
\vskip 0.15 in
\caption{
{\bf Specific heat of Ce$_2$Au$_3$In$_5$ and La$_2$Au$_3$In$_5$.}
{\bf a}, Total specific heat $c_p$ as a function of $T^3$ for Ce$_2$Au$_3$In$_5$ (filled circles) 
and La$_2$Au$_3$In$_5$ (open circles). 
{\bf b}, $c_p/T$ of La$_2$Au$_3$In$_5$ as a function of $T^2$. The solid line is a Debye-Sommerfeld fit.
}
\label{fig:CeLa235}
\end{figure}
\clearpage

\clearpage

\end{document}